\documentclass[fleqn,usenatbib]{mnras}

\usepackage{newtxtext,newtxmath}
\usepackage[T1]{fontenc}
\usepackage{ae,aecompl}

\usepackage[utf8]{inputenc}
\usepackage{graphicx}	% Including figure files
\usepackage{amsmath}	% Advanced maths commands
\usepackage{amssymb}	% Extra maths symbols
\usepackage{bm}
\usepackage{color}
\usepackage{soul}

\newcommand{\kmt}[1]{\textcolor{black}{#1}}

\title[Asteroseismology of core-collapse supernovae - I]{Towards asteroseismology of core-collapse supernovae with gravitational-wave observations - I. Cowling approximation}

\author[A.~Torres-Forn\'e, P.~Cerd\'a-Dur\'an, A.~Passamonti, and J.A.~Font]{
Alejandro Torres-Forn\'e,$^{1}$\thanks{E-mail: alejandro.torres@uv.es}
Pablo Cerd\'a-Dur\'an,$^{1}$
Andrea Passamonti,$^{2}$ 
\and and Jos\'e A. Font$^{1,3}$
\\
$^{1}$Departamento de Astronom\'ia y Astrof\'isica, Universitat de Val\`encia, c/ Dr. Moliner, 50, Burjassot (Valencia) E46100, Spain. \\
$^{2}$Departamento de F\'isica Aplicada, Universitat d'Alacant, Ap. Correos 99, E03080 Alicante, Spain. \\
$^{3}$Observatori Astron\`omic, Universitat de Val\`encia, C/ Catedr\'atico Jos\'e Beltr\'an 2, 46980, Paterna (Val\`encia), Spain.
}

\date{Accepted XXX. Received YYY; in original form ZZZ}

\pubyear{2017}

\begin{document}
\label{firstpage}
\pagerange{\pageref{firstpage}--\pageref{lastpage}}
\maketitle

\begin{abstract}
Gravitational waves from core-collapse supernovae are produced by the excitation of different oscillation modes in the proto-neutron star (PNS) and its surroundings, including the shock. In this work we study the relationship between the post-bounce oscillation spectrum of the PNS-shock system and the characteristic frequencies observed in gravitational-wave signals from core-collapse simulations. This is a fundamental first step in order to develop a procedure to infer astrophysical parameters of the PNS formed in core-collapse supernovae. Our method combines information from the oscillation spectrum of the PNS, obtained through linear-perturbation analysis in general relativity of a background physical system, with information from the gravitational-wave spectrum of the corresponding non-linear, core-collapse simulation. Using results from the simulation of the collapse of a 35 $M_{\odot}$ presupernova progenitor we show that both types of spectra are indeed related and we are able to identify the modes of oscillation of the PNS, namely g-modes, p-modes, hybrid modes, and standing-accretion-shock-instability (SASI) modes, obtaining a remarkably close correspondence with the time-frequency distribution of the gravitational-wave modes. The analysis presented in this paper provides a proof-of-concept that asteroseismology is indeed possible in the core-collapse scenario, and it may serve as a basis for future work on PNS parameter inference based on gravitational-wave observations. 
\end{abstract}


%%%%%%%%%%%%%%%%%%%%%%%%%%%%%%%%%%%%%%%%%%%%%%%%%%

\newcommand{\runi}{{\hat {\boldsymbol r}}}
\newcommand{\thetauni}{{\hat {\boldsymbol \theta}}}
\newcommand{\varphiuni}{{\hat {\boldsymbol \varphi}}}

%%%%%%%%%%%%%%%%% BODY OF PAPER %%%%%%%%%%%%%%%%%%
\section{Introduction}
%%%%%%%%%%%%%%%%%%%%%

The collapse of massive stars, i.e.~those stars with a mass larger than about 8 $M_\odot$, is among the most interesting
sources of gravitational radiation. Collapsing stars produce rich and complex waveforms, which could provide ample and important
information about the phenomenology of the scenario, specially when combined with observations of their electromagnetic emission 
and neutrino emission. The outcome of those events is either a neutron star or a black hole, typically 
(but not necessarily in the latter case) accompanied by a supernova explosion. The modelling of core collapse supernova
requires a wide variety of physical ingredients, including general relativity, a nuclear-physics-motivated equation of state (EoS), and
a detailed description of neutrino interaction \citep[see e.g.][for a review]{Janka:2007}. For sufficiently compact stellar cores
 a black hole is likely to form instead of a neutron star~\citep{OConnor:2013}, although those cases not always coincide with the most massive stars \citep[see e.g.][]{Ugliano:2012}. Recent studies have shown that, even if a black hole is formed, a successful explosion is still possible provided sufficient rotation is present in the core~\citep{Obergaulinger:2017}. In addition, it is necessary to model complex multidimensional effects and instabilities, such as convection, the standing accretion shock instability (SASI) and turbulence, which are crucial for the development of a successful supernova explosion.

The numerical modelling of the core-collapse scenario is computationally challenging and even today, with the use of the largest scientific supercomputing facilities available, we are only starting to understand the physics involved and we are probably still far away from having detailed waveforms. Unlike the binary-black-hole case, it is currently not possible to relate uniquely and unambiguously the properties of the progenitor stars (such as mass, rotation rate, metallicity, or magnetic fields) with the resulting waveforms. The reasons are the complex non-linear dynamics associated with the evolution of a fluid interacting with neutrino radiation, the stochastic and chaotic behaviour of instabilities (both during and prior to the collapse of the star), the uncertainties in stellar evolution of massive stars (specially regarding the treatment of convection, magnetic fields and angular momentum transport) and the uncertainties in the nuclear and weak interactions necessary for the EoS at high densities and neutrino radiation, respectively.

Despite the difficulties, impressive progress has been made in the last decade regarding waveforms. The core-bounce signal
is the part of the waveform which is best understood~\citep{Dimmelmeier:2002b}. Its frequency (at about 800 Hz) 
can be directly related to the rotational properties of the core~\citep{Dimmelmeier:2008, abdikamalov14, Richers:2017}. 
However, fast-rotating progenitors are not common and their bounce signal will be probably difficult to observe in typical 
slow-rotating galactic events, due to its high frequency and low amplitude. More interesting is the signal associated with the post-bounce evolution of the newly formed proto-neutron star (PNS). During that phase, the main mechanisms of gravitational waves are convection and the excitation of highly damped modes in the PNS by accreting material and instabilities (SASI). A number of groups have identified features in the gravitational-wave signal as associated with g-modes in the PNS and SASI in either 2D simulations~\citep{Murphy:2009,Mueller:2013,cerda-duran15} or 3D simulations~\citep{Kuroda:2016,Andresen:2017}. Typically the waveforms last for about $200-500$~ms until the supernova explodes \citep[see e.g][]{Mueller:2013} or, in the case of black hole formation, the typical duration is $1$~s or above \citep{cerda-duran15}. Typical frequencies raise monotonically with time due to the contraction of the PNS, whose mass steadily increases. Characteristic frequencies of the PNS can be as low as $\sim100$~Hz, specially those related to g-modes, which make them a perfect target for ground-based interferometers with the highest sensitivity at those frequencies. Two regions in the PNS appear to be susceptible to the excitation of g-modes~\citep{cerda-duran15, Andresen:2017,Kuroda:2016}, namely the surface of the PNS and the innermost cold core of the PNS. In those two regions, the specific entropy increases with the radius, resulting in convectively stable regions in which buoyancy acts as a restoring force. Moreover, quasi-radial modes in rotating cores produce a distinctive frequency-decreasing pattern in the waveforms that signals the formation of a black hole as its frequency approaches zero~\citep{cerda-duran15}. All these results imply that it may be possible to infer the properties of PNS based on the identification of mode frequencies in their waveforms, without the necessity of a complete understanding of the details about the physics involved in the core-collapse scenario.

The idea of identifying the properties of PNS based on the study of the frequencies of their normal modes of oscillation is not 
new. Studies of PNS asteroseismology have been reported in a number of works \citep[see e.g.][]{Reisenegger:1992,Ferrari:2004,Passamonti:2007, Kruger:2015, Camelio:2017}. The common approach in those cases is to study the oscillations in PNS as linear perturbations of a spherical equilibrium star (which we call ``background model'' hereafter). This results in an eigenvalue problem, whose solutions are the normal oscillation modes of the PNS. In most of the previous work a fairly simplified description of the post-bounce evolution of the PNS has been considered. Recently, \citet{Sotani:2016} have performed a linear perturbation analysis of PNS, based on simple fits to realistic 1D core-collapse simulations, to study the evolution of the mode frequencies up to several hundreds of ms after bounce. 
\citet{Fuller:2015} performed a Newtonian linear perturbation analysis using more realistic profiles from 2D simulations, but their analysis was restricted to the bounce signal of rapidly-rotating progenitors and did not consider the post-bounce evolution. So far, the presence of a standing shock above the PNS surface has not been taken into account and the oscillations have been limited to the interior of the PNS by imposing boundary conditions at the PNS surface\footnote{A notable exception is the work of~\citet{Fuller:2015} that considered the PNS surrounded by a low-density accreting region and imposed outgoing sound-wave boundary conditions in the outer boundary.}.

In this paper we present a method to perform the linear-perturbation analysis in general relativity of a background model which is the result of multi-dimensional core-collapse simulations, and includes the PNS and the hot-bubble region up to the shock location. This allows to directly  compare the frequencies of the modes obtained with our linear analysis with the actual frequencies in the gravitational-wave spectrum of the corresponding numerical simulation. This analysis provides a proof-of-concept that asteroseismology is indeed possible in the core-collapse scenario, despite the complexities of the system, and will serve as a basis for future work on PNS parameter inference based on gravitational-wave observations.

So far, attempts at parameter estimation in the context of core-collapse supernovae have been focused in the bounce signal. First steps towards waveform reconstruction and parameter estimation within this context were taken by~\citet{summerscales:2008}, followed by the work of~\citet{logue:2012} who used numerical gravitational-wave template signal to determine the supernova explosion mechanism. This pioneer work have been extended recently by \citet{powell:2016}, addressing several of the limitations  discussed in the work of~\citet{logue:2012}.

This paper is organised as follows: in Section \ref{sec:350C} we briefly describe the core-collapse model we use to compute the normal modes of oscillation and whose gravitational-wave spectrum we want to analyse. Section \ref{sec:linear_pert} describes the equations of the linear-perturbation approximation on which our study is based. In Section \ref{sec:class} we present our procedure to identify and classify the various families of oscillation modes of our physical system. In Section \ref{sec:GW} we compare the time-frequency distribution of the modes with the spectrogram of the gravitational-wave emission of the PNS and investigate their relationship. Finally in Section \ref{sec:summary} we present our conclusions and outline future extensions of this work. The paper also contains an appendix in which we test our linear-analysis method using an exact solution provided by a constant-density star. Throughout this paper we use a spacelike metric signature (-,+,+,+) and $c=G=1$ (geometrised) units, where $c$ stands for the speed of light and $G$ is Newton's gravitational constant. As customary, Greek indices run from 0 to 3, Latin indices from 1 to 3, and we use Einstein's summation convention for repeated indices.

 %%%%%%%%%%%%%%%%%%%%%
\section{Black-hole-forming model 35OC}
\label{sec:350C}
%%%%%%%%%%%%%%%%%%%%%

\begin{figure}
	\centering
		{\includegraphics[width=.49\textwidth]{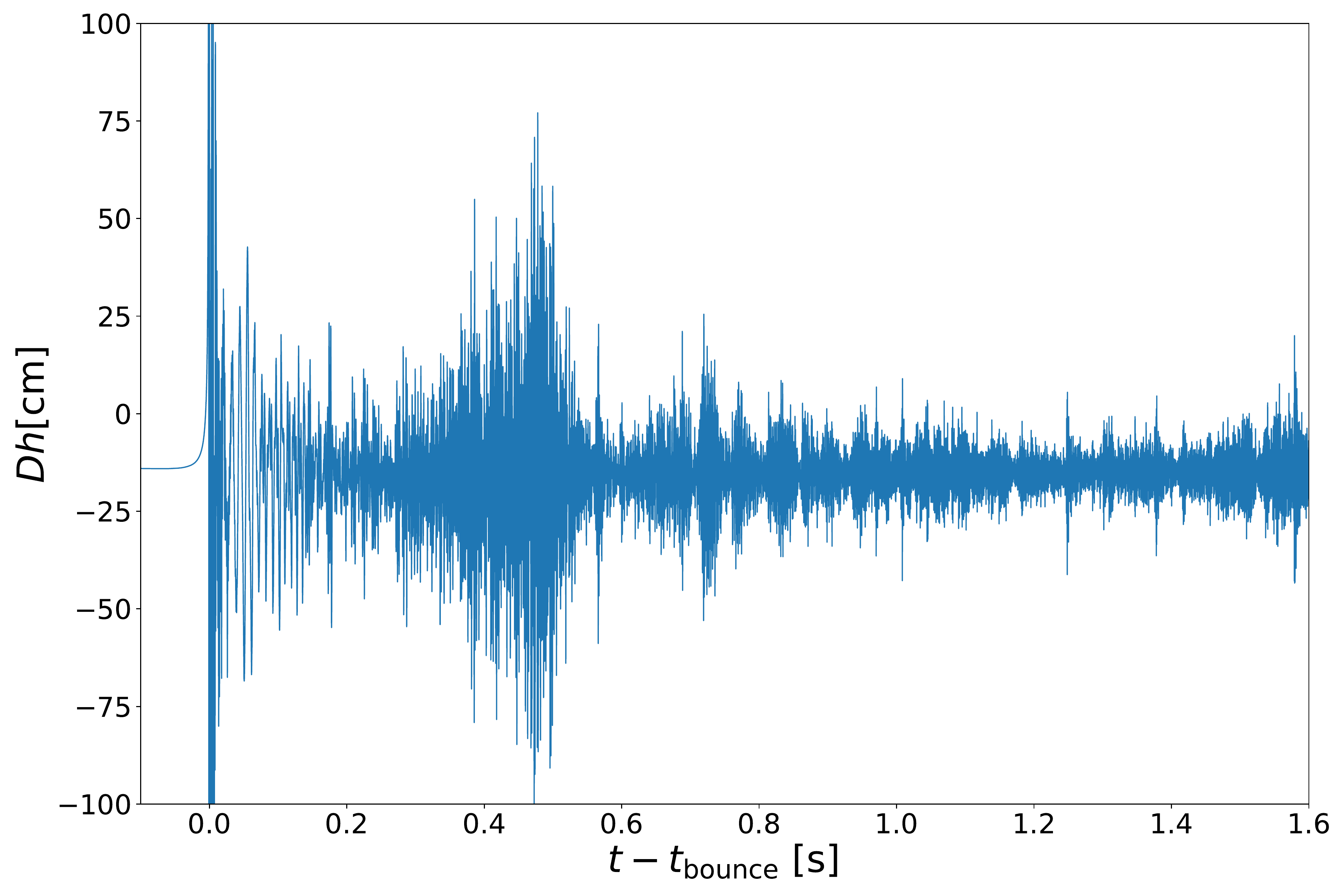}}
		\caption[Waveform of the gravitational-wave signal computed in the core-collapse simulation of~\citet{cerda-duran15}.]{Waveform of the gravitational-wave signal computed in the core-collapse simulation of~\citet{cerda-duran15} for the 35OC presupernova model at a distance $D=100$ kpc.}
		\label{fig:gw_signal}
	\end{figure}

In this work we investigate a single core-collapse model with the goal of understanding
the spectrum of eigenmodes of the coupled physical system formed by the PNS and the supernova shock wave. To accomplish this goal, we have developed  a numerical code that computes and automatically classifies the eigenmodes. We use a model we already simulated previously, so as to build confidence in our approach before proceeding to apply it to a larger set of models.
More precisely, we employ the results of the 2D core-collapse simulation performed by \citet{cerda-duran15} using the general-relativistic code {\tt CoCoNuT} \citep{dimmelmeier:2002, dimmelmeier:2005}. The progenitor is a low-metallicity $35 {\rm M}_\odot$ star at zero-age main-sequence from~\citet{woosley:2006}. This progenitor has a high rotation rate and is usually regarded as a progenitor of long-duration gamma-ray bursts (GRBs). The simulation used the LS220 EoS of~\citet{Lattimer:1991} to describe matter at high densities along with a simplified leakage scheme to approximate neutrino transport. The core of the progenitor collapses after 342.7 ms, forming a PNS and an accretion shock, after which an accretion phase ensues. The infalling matter crosses the stalled shock, heats up and falls through the hot bubble in which the fluid can dwell for some time before reaching the surface of the PNS, due to convection
and the SASI. Finally, after 1.6 s, the PNS becomes unstable to radial perturbations and collapses to a black hole. 
During this time the highly-perturbed PNS is an efficient emitter of gravitational waves, as Fig. \ref{fig:gw_signal} shows.

Throughout all the accretion phase we compute the eigenmodes of the region extending from the PNS up to the shock location.
The size of this region varies in time as the shock position changes. At post-bounce time, this region is approximately at hydrostatic equilibrium and flow velocities are small compared to the speed of sound (supersonically falling matter becomes subsonic as it crosses the shock). Therefore, we can study linear perturbations of a background provided by the result of the simulation at a given time. This approach is possible as long as the typical evolution timescales of the background are much longer than the inverse of the frequency of the modes studied. In our case, the structure of the PNS varies in a timescale of $\sim100$~ms, which limits the validity of our approach to frequencies larger than $\sim10$~Hz. Moreover, despite its relatively high rotation frequency, centrifugal forces are not dominant when compared to gravitational forces \citep[see][]{cerda-duran15}. To simplify the analysis, we perform angular averages of all quantities (weighted by the surface area of the sphere covered by each numerical cell) to compute an effective 1D model, which will be used to perform the linear perturbation analysis. Since the shock is not completely spherical in the simulation but deformed by the presence of the SASI, angular averages have to be performed with care. We first compute the averaged shock location, then rescale the radial profiles of all quantities to the average shock radius and finally perform the angular average.

\begin{figure}
	\centering
		{\includegraphics[width=.49\textwidth]{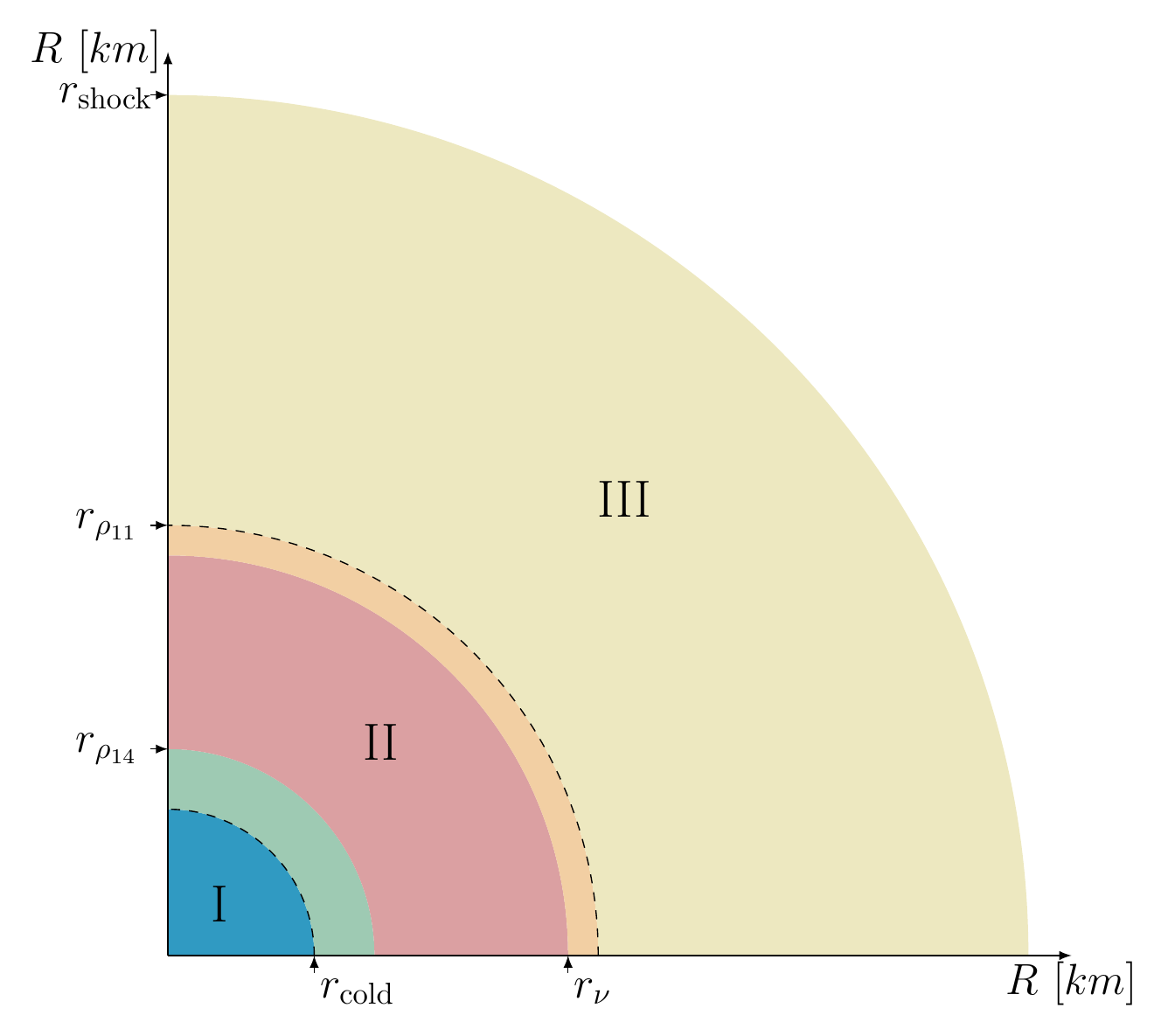}}
		\caption[Representation of the different regions of the PNS considered. ]{Representation of the different regions of the PNS considered. (See main text for details.)}
		\label{fig:regions}
	\end{figure}

For our analysis it is interesting to define three different regions between the centre of the star and the position of the accretion shock. These regions are displayed in Fig.~\ref{fig:regions}. The inner cold core (in blue) is the area between the centre and the radius at which the specific entropy is lower than $3~k_{\rm B}$ per baryon, marked in the figure as $r_{\rm cold}$. It is not easy to define the surface of the PNS. Above the cold inner core there is a broad hot mantle of $10-20$~km where the density decreases rapidly. The neutrinosphere, $r_\nu$, is typically located just below the surface and could be regarded as a proxy for the PNS radius. However, it tends to underestimate the size of the PNS slightly. We found that the point where the density becomes $10^{11}$~g~cm$^{-3}$ is a much better proxy for the PNS surface, and we use its value, $r_{\rho_{11}}$, as a definition of the 
PNS surface location in this work. Finally, the position of the shock is labeled in Fig.~\ref{fig:regions} as $r_{\rm shock}$. To summarise, the three regions are: region I, corresponding to the inner core, region II, defined as the region between the core and the surface of the PNS, corresponding to the PNS mantle, and region III, which is the area between the surface and the shock, corresponding to the hot bubble.

%
%%%%%%%%%%%%%%%%%%%%%
\section{Linear perturbations of a spherically-symmetric background}
\label{sec:linear_pert}
%%%%%%%%%%%%%%%%%%%%%

We start our analysis with the description of the perturbations of a spherically-symmetric, self-gravitating, equilibrium configuration. 
 The interested reader is addressed to \citet{kokkotas:1999} and to \citet{friedman:2013} for detailed
information on linear perturbations of compact stars and asteroseismology. Classically, this analysis was performed in Schwarzshild coordinates~\citep{Thorne:1967}. In our work we will use isotropic coordinates instead, which are closer to the gauge condition used in the original numerical simulation \citep[CFC approximation, ][]{Isenberg08,Wilson96}. Moreover, the derivation of the equations in these coordinates also bears resemblance with the equations in the Newtonian case \citep[see][]{Reisenegger:1992}, which makes it easier to identify the role of the different terms in the equations and interpret the solutions. This choice of gauge also makes it straightforward to perform the mode analysis of Newtonian simulations.

Let us consider a $3+1$ foliation of spacetime in coordinates $(t,x^i)$, in which the metric can be written as
\begin{equation}
ds^2 = g_{\mu\nu}dx^\mu dx^\nu = (\beta^i\beta_i-\alpha^2) dt^2 + 2 \beta_i dt dx^i + \gamma_{ij} dx^i dx^j,
\end{equation}
where $\beta^i$ is the shift 3-vector, $\alpha$ is the lapse function, and $\gamma_{ij}$ is the spatial 3-metric. In isotropic coordinates, 
the background metric of a static and spherically-symmetric configuration can be written as
\begin{equation}
ds^2 = g_{\mu\nu}dx^\mu dx^\nu = -\alpha^2 dt^2 + \psi^4 f_{ij} dx^i dx^j,
\end{equation}
where $\psi$ is the conformal factor and $f_{ij}$ is the flat spatial 3-metric. In the spherically-symmetric and static limit, Einstein's equations for the CFC 
metric\footnote{Note that in spherical symmetry, the CFC metric is no longer an approximation and it is equivalent to choosing maximal slicing and isotropic coordinates.} read
\begin{align}
\Delta \psi &= - 2 \pi \psi^{5} E, \label{eq:cfc:hamiltonian}\\
\Delta (\alpha \psi) &= 2\pi \alpha \psi^{5} \left [ E + 2 S \right ], \label{eq:cfc:gauge}
\end{align}
where $\Delta$ is the Laplacian operator with respect to the flat 3-metric. In this case $\beta_i =0$ 
and $\gamma_{ij} = \psi^4 f_{ij}$. The energy-momentum content couples
to the spacetime geometry through the projections of the energy-momentum tensor, $T_{\mu\nu}$, onto the $3+1$ foliation
\begin{align}
E &\equiv \alpha^2 T^{00}, &\quad
S_i &\equiv - (T_{0i} - T_{ij} \beta^j), \\
S_{ij} & \equiv T_{ij}, &\quad
S & \equiv S_{ij} \gamma^{ij}.
\end{align}
We consider a perfect fluid, for which the energy-momentum tensor is given by
\begin{equation}
T^{\mu\nu} = \rho h u^{\mu} u^{\nu} + P g^{\mu\nu} \,,
\end{equation}
where $\rho$ is the rest-mass density, $P$ is the pressure, $u^{\mu}$ is the 4-velocity,
$h\equiv 1 + \epsilon + P/\rho$ is the specific enthalpy, and $\epsilon$ is the specific internal
energy. It is useful to define the energy density as $e \equiv \rho (1 + \epsilon)$.

If we consider the Bianchi identities and the conservation of the number of baryons (continuity equation) the equations of general relativistic hydrodynamics w.r.t.~a coordinate basis read~\citep{Banyuls:1996}:
\begin{equation}
\frac{1}{\sqrt{\gamma}} \partial_t \left [ \sqrt{\gamma}D \right ] + \frac{1}{\sqrt{\gamma}} \partial_i \left [ \sqrt{\gamma} D v^{*i} \right]
 = 0, \label{eq:continuity} 
 \end{equation}
 \begin{equation}
\frac{1}{\sqrt{\gamma}} \partial_t \left [ \sqrt{\gamma} S_j \right ] 
+ \frac{1}{\sqrt{\gamma}} \partial_i \left [ \sqrt{\gamma} S_j v^{*i} \right ]
+ \alpha \partial_i P 
= \frac{\alpha \rho h}{2} u^{\mu}u^{\nu} \partial_j g_{\mu\nu}, \label{eq:momentum} 
\end{equation}
\begin{align}
\frac{1}{\sqrt{\gamma}} \partial_t \left [ \sqrt{\gamma} E \right ]
+ \frac{1}{\sqrt{\gamma}} &\nabla_i \left [ \sqrt{\gamma} \left ( E v^{*i} + \alpha P v^i \right) \right ] = \nonumber\\
&= \alpha^2 \left(
T^{\mu0} \partial_\mu \ln \alpha - T^{\mu\nu} \Gamma^0_{\mu\nu}
\right ), \label{eq:energy}
\end{align}
where $\gamma$ is the determinant of the three-metric, $\Gamma^{\lambda}_{\mu\nu}$ is the Christoffel symbol associated with the four-metric, and the conserved quantities are defined as
\begin{equation}
D = \rho W, \qquad S_j = \rho h W^2 v_j , \qquad
 E = \rho h W^2 - P\,,
\end{equation}
where $W=1/\sqrt{1-v_iv^i}$ is the Lorentz factor.
The Eulerian and ``advective'' velocities are, respectively, 
\begin{align}
v^i & = \frac{u^i}{W} +\frac{\beta^i}{\alpha},\quad & \quad v^{*i} &= \frac{u^i}{u^0} =\alpha v^i - \beta^i.
\end{align}

Let us consider a solution of the hydrodynamics equations that is in equilibrium ($\partial_t=0$) and is static ($v^i=0$). 
In this case Eq.~(\ref{eq:momentum}) reads
\begin{equation}
\frac{1}{\rho h} \partial_i P = -\partial_i \ln \alpha \equiv G_i, \label{eq:G}
\end{equation}
where $G_i$ is the gravitational acceleration, in the Newtonian limit, whose only non-zero component is $G_r\equiv G$. The solution of Eq.~(\ref{eq:G}) corresponds to the unperturbed state or background solution.

Next, we consider linear adiabatic perturbations of the hydrodynamics equations with respect to the background equilibrium 
configuration. We denote Eulerian perturbations of the different quantities with $\delta$, e.g. for the rest-mass density $\rho$, the
Eulerian perturbation is $\delta \rho$. The linearised equations are obtained by substituting $\rho \to \rho + \delta\rho$, and so forth,
in Eqs.~(\ref{eq:continuity}-\ref{eq:energy}). Note that $\rho$, $P$, etc, correspond to the background value in the linearised equations.
Furthermore, we consider the Cowling approximation, i.e. we do not take into account perturbations of the metric ($\delta\alpha =\delta\psi=\delta \beta^j=0$). We discuss below the impact of this approximation in our results.

We denote as $\xi^i$ the Lagrangian displacement of a fluid element with respect to its position at rest. Its value is related
to the advective velocity as
\begin{equation}
\partial_t \xi^i = \delta v^{*i}.
\end{equation}
The Lagrangian perturbation of any quantity, e.g. $\rho$, is related to the Eulerian perturbations as
\begin{equation}
\Delta \rho = \delta \rho + \xi^i\partial_i \rho.
\end{equation}
The linearised version of Eqs.~(\ref{eq:continuity}) and (\ref{eq:momentum}) are
\begin{align}
\frac{ \Delta \rho }{\rho} &= - \left( \partial_i \xi^i + \xi^i \partial_i \ln \sqrt{\gamma} \right) \, , \label{eq:cnt} \\
\rho h \, \partial_t \delta v_i + \alpha \partial_i \delta P &= - \delta \left( \rho h \right) \partial_i \alpha \,, \label{eq:meq2}
\end{align}
where $\sqrt{\gamma}$ is the determinant of the 3-metric. We use spherical coordinates, $\{r,\theta,\varphi\}$, in which $\sqrt{\gamma} = \psi^6 r^2 \sin\theta$.
Since we are considering adiabatic perturbations, Eq.~(\ref{eq:energy}) does not add additional information to the problem.

The condition of adiabaticity of the perturbations implies that 
\begin{equation}
\frac{\Delta P}{\Delta \rho} = \left . \frac{\partial P}{ \partial \rho} \right |_{\rm adiabatic} = h c^2_{{\rm s}} = \frac{P}{\rho}\Gamma_1,
\end{equation}
where $c_{\rm s}$ is the relativistic speed of sound and $\Gamma_1$ is the adiabatic index.
This allows us to write
\begin{equation}
\delta \left( \rho h \right) = \left( 1 + \frac{1}{c_s^2} \right) \delta P - \rho h \, \xi^i \mathcal{B}_{i} \, , \label{eq:rh}
\end{equation} 
where
\begin{equation}
\mathcal{B}_i \equiv \frac{\partial_i e}{\rho h} - \frac{1}{\Gamma_1}\frac{\partial_i P}{P},
\end{equation}
is the relativistic version of the Schwarzschild discriminant. Since the background is spherically symmetric, 
the only non-zero component is ${\mathcal B}_r\equiv {\mathcal B}$.

The radial and angular parts of equation (\ref{eq:meq2}) are given by 
\begin{align}
& \rho h \, \psi^4 \alpha^{-2} \frac{ \partial^2 \xi^r }{\partial t^2 } + \partial_r \delta P = \delta \left( \rho h \right) G ,  \label{eq:vr} \\
& \rho h \, \psi^4 \alpha^{-2} r^2 \frac{ \partial^2 \xi^{\theta} }{\partial t^2 } + \partial_{\theta} \delta P = 0 , \label{eq:vth} \\
& \rho h \, \psi^4 \alpha^{-2} r^2 \sin^2\theta\frac{ \partial^2 \xi^{\varphi} }{\partial t^2 } + \partial_{\varphi} \delta P = 0, \label{eq:vph} 
\end{align} 
where we have used that, in the coordinate basis, the covariant components of the velocity are given by $\delta v_r = \psi^4 \delta v^r$, $\delta v_{\theta} = r^2 \psi^4 \delta v^{\theta}$
and $\delta v_{\varphi} = r^2 \sin^2 {\theta} \psi^4 \delta v^{\varphi}$.

We perform an expansion of the perturbations with a harmonic time dependence of frequency $\sigma$ and a spherical-harmonic 
expansion for the angular dependence
\begin{align}
\delta P = \delta \hat{P} \,\,& Y_{lm} e^{- i \sigma t}, \\
{\bf \xi}^r = \eta_r \,\, &Y_{lm} e^{- i \sigma t} , \\
{\bf \xi}^\theta = \eta_\perp \,\,& \frac{1}{r^{2}} \partial_\theta Y_{lm} e^{- i \sigma t}, \\
{\bf \xi}^\varphi = \eta_\perp \,\,& \frac{1}{r^2 \sin^2\theta} \partial_\varphi Y_{lm} e^{- i \sigma t}. 
\end{align}
Here we only consider polar perturbations,  because purely fluid non-rotating stars do not sustain axial (torsional) oscillations.
The quantities $\eta_r, \eta_{\perp}$ and the scalar perturbations with the hat, i.e. $\delta \hat P$, are only function 
of the radial coordinate. For $l\neq0$, by inserting the spherical-harmonic expansion into equations (\ref{eq:vr})-(\ref{eq:vth}) we obtain:
\begin{align}
& - \sigma^2 \rho h \, \psi^4 \alpha^{-2} \eta_r + \partial_r \delta \hat P = \delta \left( \rho h \right) G, \label{eq:vr2} \\
& - \sigma^2 \rho h \, \psi^4 \alpha^{-2} \eta_{\perp} + \delta \hat P = 0. \label{eq:vth2} 
\end{align} 
$l=0$ modes of spherical stars do not emit gravitational waves and hence are not considered in this work.
From Eq.~(\ref{eq:vth2}) we get
\begin{equation}
\delta \hat P = q \sigma^2 \eta_\perp \, , \label{eq:dp}
\end{equation}
where for convenience we have defined $q \equiv \rho h \, \alpha^{-2} \psi^4$. 

Using Eqs.~(\ref{eq:dp}) and (\ref{eq:rh}) to simplify Eqs.~(\ref{eq:cnt}) and (\ref{eq:vr2}) we obtain
\begin{align}
\partial_r \eta_{r} 
+ \left[ \frac{2}{r} + \frac{1}{\Gamma_1}\frac{ \partial_r P }{P } + 6 \frac{ \partial_r \psi }{\psi}  \right] \eta_r 
+ \frac{\psi^4 }{\alpha^2 c^2_s} \left( \sigma^2 - \mathcal{L}^2 \right) \eta_\perp = 0 \, , \label{eq:er} \\
\partial_r \eta_\perp 
- \left( 1 - \frac{ \mathcal{N}^2 }{\sigma^2} \right) \eta_r 
+ \left[ \partial_r \ln q - G \left( 1 + \frac{1}{c^2_s} \right) \right] \eta_\perp =0,
\label{eq:eth} 
\end{align}
where $ \mathcal{N}$ is the relativistic Brunt-V\"ais\"al\"a frequency defined as 
\begin{equation}
 \mathcal{N}^2 \equiv \frac{ \alpha^2 }{\psi^4 } G^i \mathcal{B}_i = \frac{\alpha^2}{\psi^{4}} \mathcal{B} G \, ,
\end{equation}
and $\mathcal{L}$ is the relativistic Lamb frequency defined as
\begin{equation}
\mathcal{L}^2 \equiv \frac{ \alpha^2 }{\psi^4} c^2_s \frac{l\left(l+1\right)}{r^2} \, .
\end{equation}
To simplify the discussion we define the coefficients $A$, $B$, $C$ and $D$ in Eqs.~(\ref{eq:er}) and
(\ref{eq:eth}) such that the equations take the form
\begin{align}
\partial_r \eta_r &= A \eta_r + B \eta_{\perp}, \\
\partial_r \eta_\perp &= C \eta_r + D \eta_{\perp}. 
\end{align}
\subsection{Plane-wave limit}
In the plane-wave approximation $\eta_r, \eta_\perp \sim e^{ikr}$, we can write the following dispersion relation from 
the last two equations
\begin{equation}
 -k^2 - i k (A+D) + AD - BC = 0.
\end{equation} 
The real part of $k$ is non zero, i.e.~locally there exist plane-wave solutions, if and only if 
\begin{equation}
4 B C < (A-D)^2.
\end{equation}
Therefore, a sufficient condition to have a real part of $k$ is that $BC<0$ and hence
\begin{equation}
 (\sigma^2 - \mathcal{L}^2) (\sigma^2 - \mathcal{N}^2) > 0 \quad \to\quad \mathrm{Re}(k)\ne 0.
\end{equation}
In cold NSs, typically $\mathcal{L}^2 >> \mathcal{N}^2 $, thus the solutions can be either $ \sigma^2 > \mathcal{L}^2 > \mathcal{N}^2 $ (acoustic modes) or 
$ \sigma^2 < \mathcal{N}^2 < \mathcal{L}^2 $ (g-modes). However, in the core-collapse scenario considered in this work we deal with a new-born PNS. Therefore, the PNS is still hot and surrounded by an extended mantle and those simplifying assumptions do not hold.

%%%%%%%%%%%%%%%%%%%%%
\subsection{G-mode limit}
\label{sec:gmodes}
%%%%%%%%%%%%%%%%%%%%%

G-modes (or gravity modes) can appear in regions of the star where buoyancy acts as a restoring force. In those regions the Brunt-V\"ais\"al\"a frequency
is such that $\mathcal{N}^2>0$, i.e.~the regions are stable to convective instabilities. It is possible to obtain approximate solutions for 
the g-modes in a star by neglecting sound waves in the system. The calculation of g-modes in this approximation can be used 
to have a rough idea of what are the typical g-mode frequencies of PNSs. It also serves as a basis to identify which modes belong to the g-mode class when the full analysis with no approximations is performed. It is possible to remove acoustic waves from the system by taking the limit of Eqs.~(\ref{eq:er}) and (\ref{eq:eth}) when the speed of sound tends to infinity, $c_s^2 \rightarrow \infty $. 
In this limit the equations read
\begin{align}
\partial_r \eta_{r} 
+ \left[ \frac{2}{r} + 6 \frac{ \partial_r \psi }{\psi}  \right] \eta_r 
- \frac{l(l+1)}{r^2} \eta_\perp = 0 \, , \label{eq:g_er} \\
\partial_r \eta_\perp 
- \left( 1 - \frac{ \mathcal{N}^2 }{\sigma^2} \right) \eta_r 
+ \left[ \partial_r \ln q - G \right] \eta_\perp =0.
\label{eq:g_eth} 
\end{align}
%

%%%%%%%%%%%%%%%%%%%%%
\subsection{Acoustic-mode limit}
\label{sec:pmodes}
%%%%%%%%%%%%%%%%%%%%%

Similarly to the previous case, it is also possible to remove buoyancy from the system by considering only modes 
supported by sound waves, i.e.~p-modes. The acoustic limit is reached by setting $\mathcal{B}=0$, and therefore, 
$\mathcal{N}^2=0$. The resulting system of equations reads,
\begin{align}
\partial_r \eta_{r} 
+ \left[ \frac{2}{r} + \frac{ \partial_r e}{\rho h } + 6 \frac{ \partial_r \psi }{\psi}  \right] \eta_r 
+ \frac{\psi^4 }{\alpha^2 c^2_s} \left( \sigma^2 - \mathcal{L}^2 \right) \eta_\perp = 0 \, , \label{eq:p_er} \\
\partial_r \eta_\perp 
- \eta_r 
+ \left[ \partial_r \ln q - G \left( 1 + \frac{1}{c^2_s} \right) \right] \eta_\perp =0.
\label{eq:p_eth} 
\end{align}

%%%%%%%%%%%%%%%%%%%%%
\subsection{Boundary conditions}
%%%%%%%%%%%%%%%%%%%%%

We impose boundary conditions at the shock location. The shock is a sonic point at which the flow decelerates from a 
supersonic regime (outside) to a subsonic regime (inside). Therefore, all the characteristic curves in the inner part of the shock,
where we need to impose boundary conditions, are pointing inwards; a wave propagating outwards from the inside will stall
when reaching the shock location. In other words, any (radial) perturbation at the shock location should have zero displacement
\begin{equation}
\xi_r|_\textrm{shock} = 0 \quad \to \quad \eta_r|_\textrm{shock} = 0.
\end{equation}
At the origin ($r=0$) it is sufficient to impose regularity \citep[see][]{Reisenegger:1992}
\begin{equation}
\eta_r|_{r=0} = l \, \eta_\perp|_{r_0} \propto r^{l-1}. \label{eq:regularity}
\end{equation}

%%%%%%%%%%
\subsection{Eigenmode computation}
%%%%%%%%%%

The procedure to compute the eigenmodes (``modes'' hereafter) entails the integration of Eqs.~(\ref{eq:er}) and (\ref{eq:eth}) from the centre of the PNS to the location of the shock for different values of $\sigma$. \kmt{We use the same radial grid of the core-collapse simulation of~\cite{cerda-duran15} to perform the mode computation.}
The specification of the initial data at the origin is somewhat arbitrary as long as it fulfils Eq.~(\ref{eq:regularity}). Those values of $\sigma$ such that $\eta_r$ vanishes at the shock correspond to eigenvalues of the system. Accurate eigenvalues can be obtained by finding the roots of $\eta_r|_\textrm{shock} (\sigma) =0$ (e.g.~using the bisection method), where $\eta_r|_\textrm{shock} (\sigma)$ is the value of $\eta_r|_\textrm{shock}$ after integrating for a given value of $\sigma$. The integration of Eqs.~(\ref{eq:er}) and (\ref{eq:eth}) is performed by means of the trapezoidal rule, which is second-order accurate and ensures the stability of the integration due to its implicit character.  \kmt{Indeed, our numerical procedure converges with the expected second-order accuracy when increasing the resolution. 
Namely, we have checked that when doubling the number of points in the integration the eigenfunction amplitude does not vary by more than 10\%  and the eigenmode frequency by no more than $1\%$.  Further}  
assessment of this numerical procedure to determine the eigenvalues and eigenmodes is shown in Appendix~\ref{sec:appendix} using  a simple test with known analytical solution. 
The same numerical procedure is employed when computing the approximate g-modes, Eqs.~(\ref{eq:g_er}) and (\ref{eq:g_eth}), and the approximate p-modes, Eqs.~(\ref{eq:p_er}) and (\ref{eq:p_eth}). Hereafter, we refer as ``approximate modes'' to the solutions of the latter two approximate systems, to distinguish them from the ``complete modes'' (usually just referred as ``modes'') resulting from the complete system of equations.
	\begin{figure}
	\centering
		{\includegraphics[width=.49\textwidth]{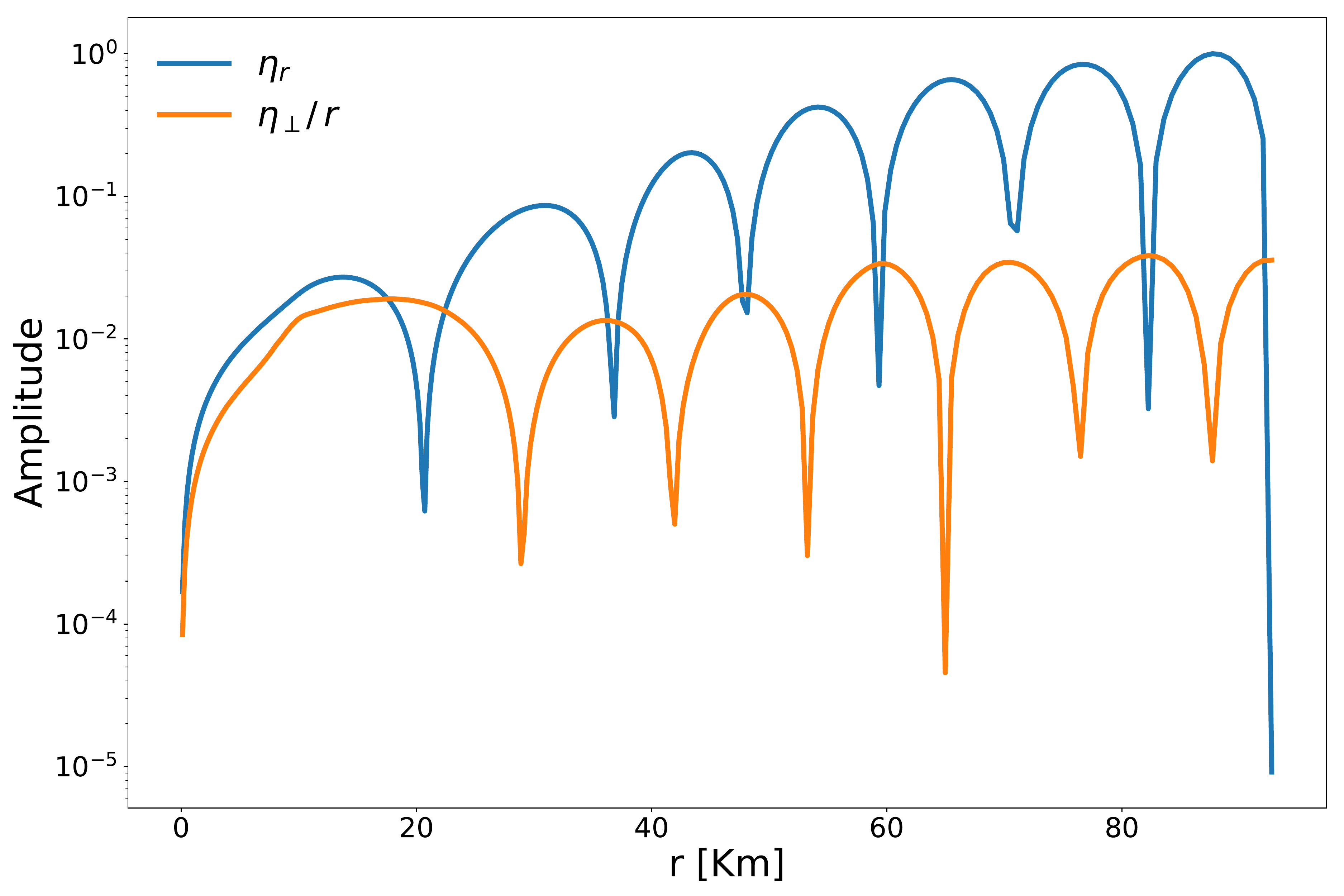}}
		\caption[Example of the radial and perpendicular components of a mode.]{Absolute values of the radial component $\eta_r$ (blue line) and of the perpendicular component $\eta_\perp$ (orange) of one example mode computed $\approx 1$ s after core bounce. The mode frequency is 2608 Hz. The $y$-axis represents the normalised amplitude in logarithmic scale, while the $x$-axis is the radius in km which ends at the shock. }
		\label{fig:mode_example}
	\end{figure}
Fig.~\ref{fig:mode_example} shows an example of the absolute value of both, the radial and the perpendicular components of the eigenfunction corresponding to a frequency of 2608 Hz calculated 1 s after core bounce. In this case both components have several nodes inside the domain. The number of nodes is used as a basis for the identification and classification of the different modes, as we show in the next sections.

%%%%%%%%%%%%%%%%%%%%%
\subsection{Energy and radiated power}
%%%%%%%%%%%%%%%%%%%%%

In the Newtonian limit, the energy stored in a mode with a certain amplitude can be approximated as
\begin{equation}
E= \frac{\sigma^2}{2}\int_0^{r_{\rm shock}}\rho~r^2 \left[\eta_r^2+l(l+1)\frac{\eta_\perp^2}{r^2} \right]dr\,.
\label{eq:energy2}
\end{equation}
The integrand can be identified as the energy density $\mathcal{E}$.
Following~\citet{thorne:1969} it is possible to compute the total radiated power
in gravitational waves by each mode with $l\ge 2$,
\begin{equation}
P_g = \frac{1}{8\pi}\frac{(l+1)(l+2)}{(l-1)l}\left[
\frac{4\pi \sigma^{l+1}}{(2l+1)!!}\int_0^{r_{\rm shock}}~\delta{\hat\rho}~r^{l+2} dr\right]^2,
\label{eq:power}
\end{equation}
where  
\begin{equation}
\delta{\hat\rho} \approx \rho \left ( 
\frac{\mathcal{N}^2}{G}\eta_r + \frac{\sigma^2}{c_s^2}\eta_\perp \right ).
\label{eq:rho1}
\end{equation}

%%%%%%%%%%%%%%%
\section{Mode classification}
\label{sec:class}
%%%%%%%%%%%%%%%

In addition to the simple classification of the modes as a function of the frequency, our goal is to classify the modes in a way which allows
us to identify the contribution of each mode to the gravitational-wave emission. In this section we focus on the $l=2, m=0$ modes. Modes for other values of $l>2$ look qualitatively similar and will be discussed in the next section. For non-rotating and non-magnetized stars, non-axisymmetric modes with $m\ne0$, have the same frequency as modes with $m=0$, for the same value of $l$, and are not considered.

%%%%%%%%%%%%%%%%%%%%%%
\subsection{Number of nodes}
%%%%%%%%%%%%%%%%%%%%%%

Our first attempt to classify the modes is according to their number of nodes. We define the number of nodes 
as the number of sign changes of the radial function $\eta_r$. However, in regions where the value of $\eta_r$ is small, 
numerical discretisation errors of the equations can induce small fluctuations of the eigenfunction that may lead to a miscount
of the number of nodes. Therefore, we try to minimise as much as possible this error in our counting algorithm; for example we do not
count nodes as different unless they are more than a few numerical cells apart, to avoid possible high-frequency numerical noise.
However, there could still be node miscounts in a few cases. While this issue could be solved by counting the number of nodes
manually, our aim is to build a fully automatised, and reliable, algorithm for the mode classification.
	\begin{figure}
	\centering
		{\includegraphics[width=.49\textwidth]{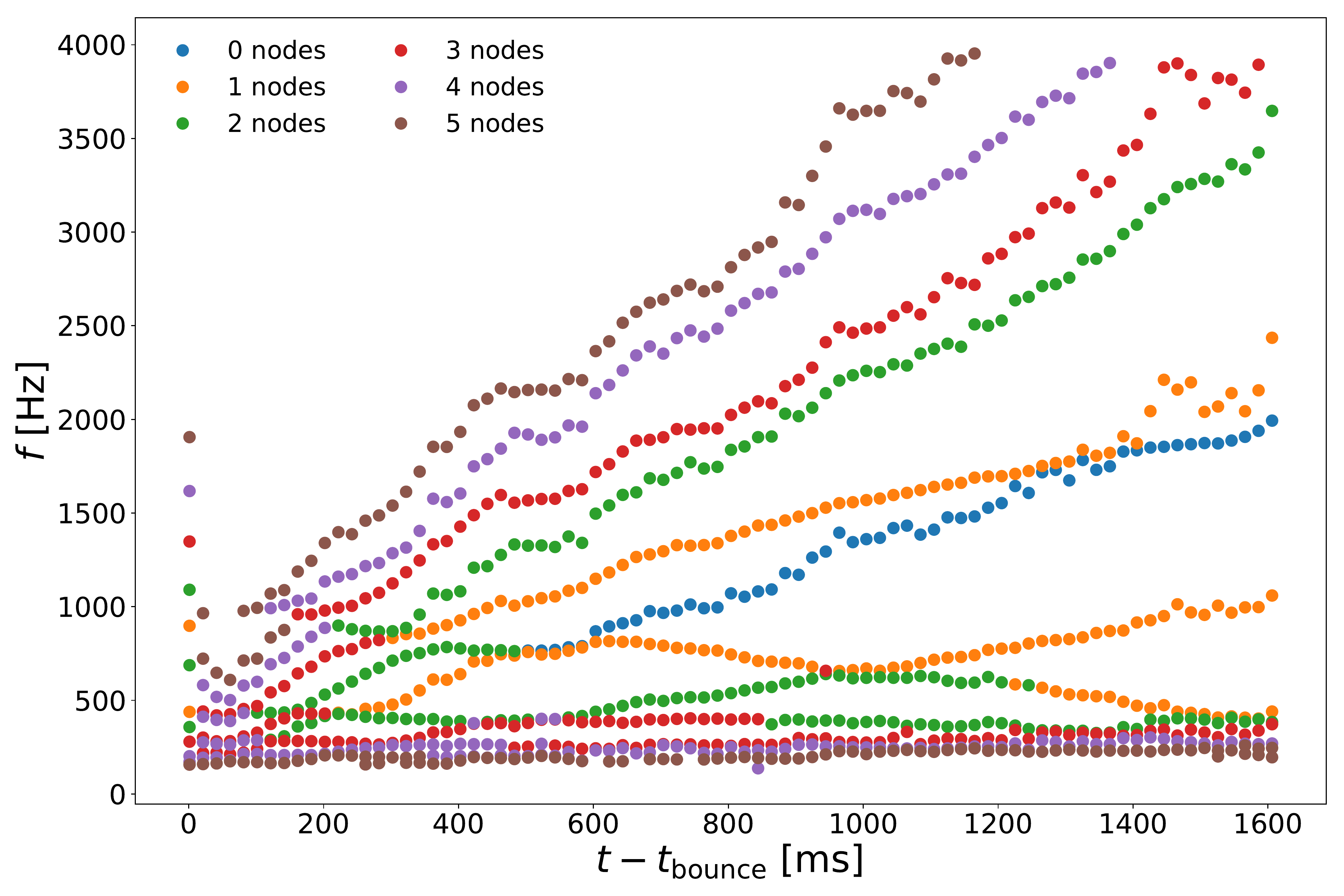}}
		\caption[Time-frequency diagram of the modes.]{Post-bounce time evolution of the frequency of the eigenmodes
		of the system. Modes with the same number of nodes are represented with the same colour. 
		Only modes with less than five nodes are represented.}
		\label{fig:tf-nnodes}
	\end{figure}

Fig.~\ref{fig:tf-nnodes} shows the post-bounce evolution of the mode frequencies, with modes classified in different colours according to their number of nodes. Note that other classification methods are possible, as those based on the continuity of mode frequency or similar features. Some interesting conclusions can be extracted from this figure. Our simple classification in terms of the number of nodes allows to tell modes apart and to follow the temporal evolution of their frequencies during the
simulation.  After $\sim 200$ ms, at which time the accretion rate drops~\citep[see] [for details]{cerda-duran15} different types of modes start to split away, and it becomes clear that there is a class of modes with a rapidly-increasing frequency and another class in which the frequency barely changes with time or even decreases. As we show in the next sections, these two classes correspond approximately to p-modes and g-modes, respectively. The jumps in frequency between consecutive times are due to changes in the accretion-shock position, which is mainly due to the SASI. We call fundamental mode (or f-mode), $^lf$, the mode with zero radial nodes. In Fig.~\ref{fig:tf-nnodes} this mode is shown in blue and is clearly distinguishable starting from $\sim 800$ ms, with a frequency of around 1000 Hz. Once the fundamental mode is identified it becomes apparent that it separates two classes of modes: one class above the f-mode, in which the number of nodes increases with increasing frequency and another class of modes below the f-mode in which the number of nodes increases with decreasing frequency. This is the expected behaviour of p-modes and g-modes, respectively. Before $\sim 800$ ms there is no mode with zero radial nodes to be identified as f-mode. The two classes of modes seem to mix and cross. At each crossing there is a change on the number of nodes. This behaviour, the so-called avoided crossing of frequencies, is typical in linear analysis of oscillations, when modes of different nature (in this case p-modes and g-modes) have similar frequencies~\citep[see e.g.][]{Stergioulas:2003}. 

%%%%%%%%%%%%%%%%%%%%%%
\subsection{Mode identification}
%%%%%%%%%%%%%%%%%%%%%%

\begin{figure}
	\centering
		{\includegraphics[width=.49\textwidth]{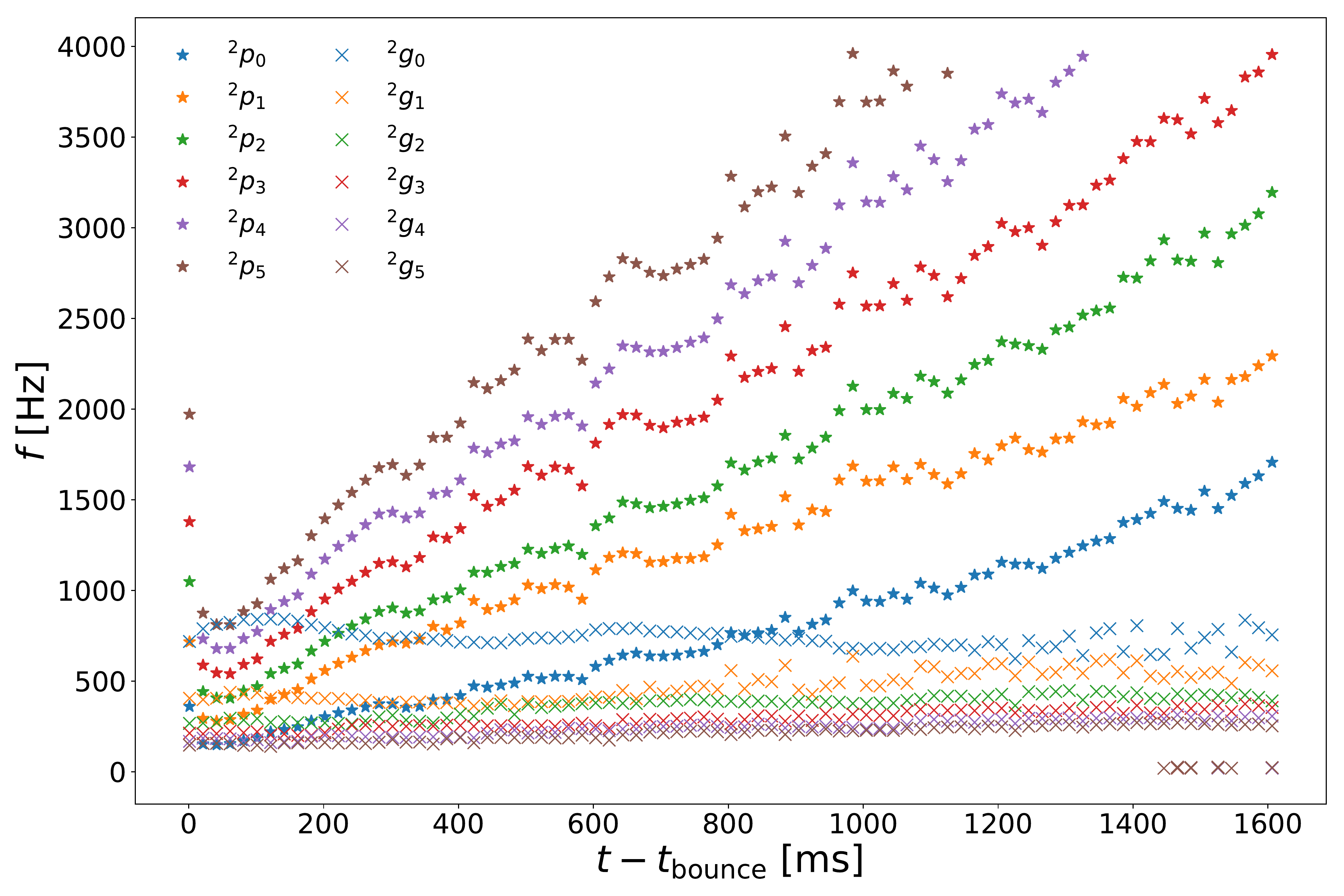}}
		\caption[Time-frequency comparison between approximate g-modes and p-modes.]
		{Time-frequency diagram of the approximate modes computed in the g-mode limit (crosses) 
		and in the p-mode limit (stars). Modes with the same number of nodes are represented with the same colour. 
		Only modes with less than five nodes are represented.}
		\label{fig:tf-gp_modes}
	\end{figure}

The previous analysis suggests that in the evolution of the new-born PNS there exist at least two types of modes. To check if these classes correspond to the theoretical separation in g-modes and acoustic-modes (p-modes), we compute them in the approximations presented in Sections~\ref{sec:gmodes} and~\ref{sec:pmodes} and calculate the number of nodes in each case. We label them as $^lg_n$ and $^lp_n$, respectively, with index $n$ being the number of nodes. Fig.~\ref{fig:tf-gp_modes} shows the results of both approximate g-modes (crosses) and p-modes (stars). It is clear that for the approximate g-modes the number of nodes increases as the frequency decreases. In contrast, for the approximate p-modes this relation is the opposite, increasing the number of nodes with increasing frequency. This behaviour supports the identification hypothesis we have done of the complete modes in the previous section.

We turn next to analyse the time evolution of the frequency of both classes of approximate modes. The fundamental approximate g-mode, labelled as $^2g_0$, has a frequency in the range $f\in[500,1000]$ Hz, varying non-monotonically but slowly between the hydrodynamical bounce and the black hole formation at the end of the simulation. The rest of g-modes have similar evolution albeit at lower frequencies. This is the expected behaviour of g-modes living inside the PNS, whose frequencies trace the surface gravity 
of the region generating the modes. The fundamental approximate p-mode $^2p_0$ is the one with the lowest frequency. As the system evolves all approximate p-mode frequencies increase almost monotonically. This kind of modes appear as standing sound waves trapped between the surface of the PNS and the shock. 
However, since the PNS surface is an extended region and not a hard wall, sound waves are able to penetrate into the PNS, so the eigenmode frequency and structure 
may not only be determined by the structure of the cavity but also by the inner structure of the PNS.
As the shock contracts during the evolution, the radial extent of this cavity decreases and the frequency of p-modes increases. 

	\begin{figure}
	\centering
		{\includegraphics[width=.49\textwidth]{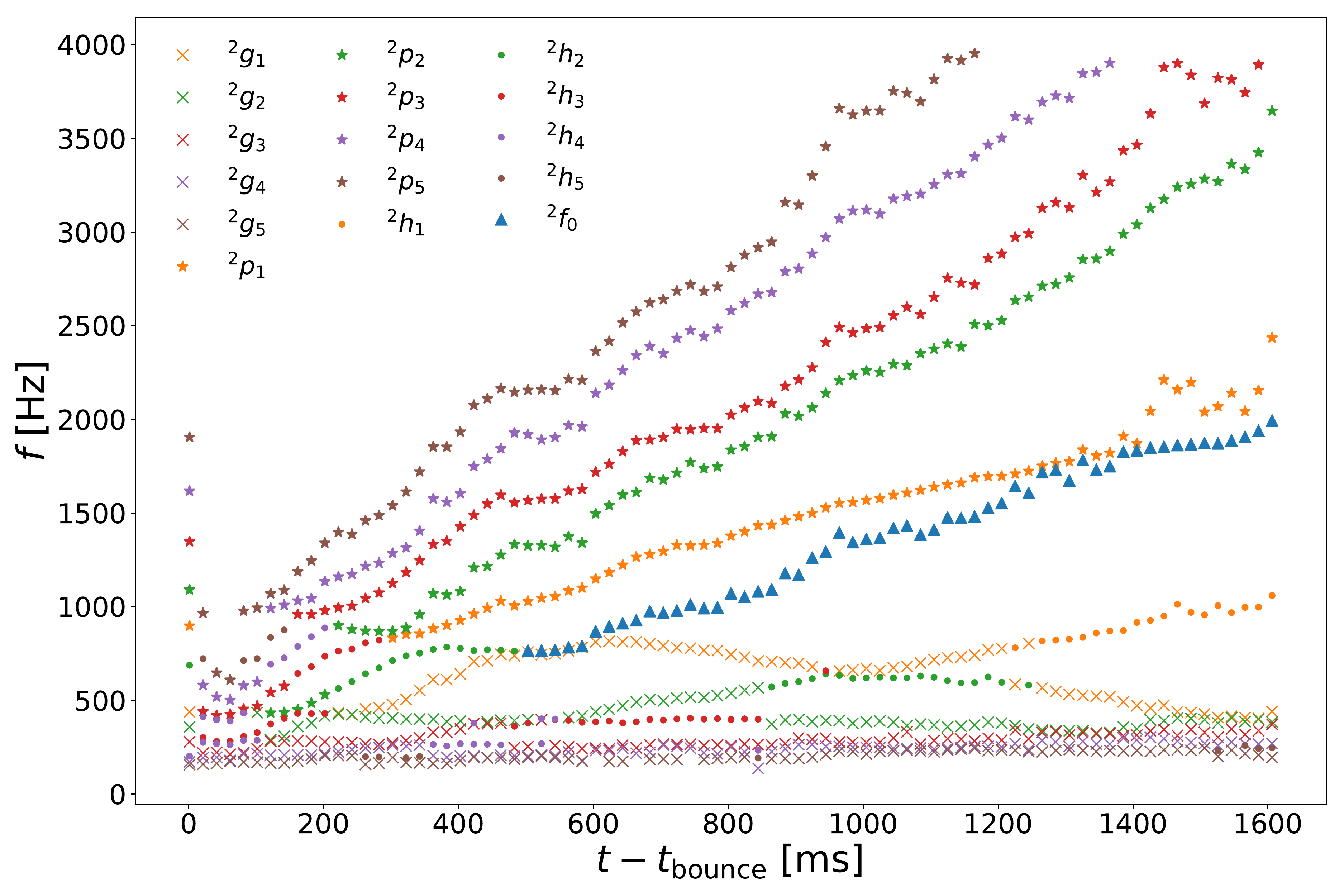}}
		\caption[Time-frequency diagram of all (complete) modes separated in classes.]{Time-frequency diagram of the
		 modes of the complete problem separated in g-modes (crosses), p-modes (stars), f-mode (triangles) and h-modes (dots). 
		 Modes with the same number of nodes are represented with the same colour. Only modes with less than five nodes are represented. }
		\label{fig:tf-classes} 
	\end{figure}

In view of these results, we have devised a method for classifying the complete modes as p-modes or g-modes: we define the $\rm{n}^{\rm th}$
p-mode/g-mode, for a given time, as the mode with the largest/lowest frequency with n nodes. Those modes not falling in either category 
are classified as hybrid modes (h-modes, hereafter). The result of this classification is shown in Fig.~\ref{fig:tf-classes}. It is interesting to compare this figure with Fig.~\ref{fig:tf-gp_modes}. There is clearly a good match between the classification procedure of the complete eigenvalue problem and the frequencies of the approximate g-modes and p-modes calculations, specially at frequencies well above or below the fundamental-mode frequency. This makes us confident that our procedure is accurately classifying different modes. We also note that a careful look to both figures shows a tendency for complete modes to have somewhat larger frequencies than their corresponding approximate modes. As expected, hybrid modes appear mainly at the crossing of g-modes and p-modes, although some may appear at different frequencies and may persist for the entire evolution (see Fig.~\ref{fig:tf-classes}).

\begin{figure}
	\centering
		{\includegraphics[width=.49\textwidth]{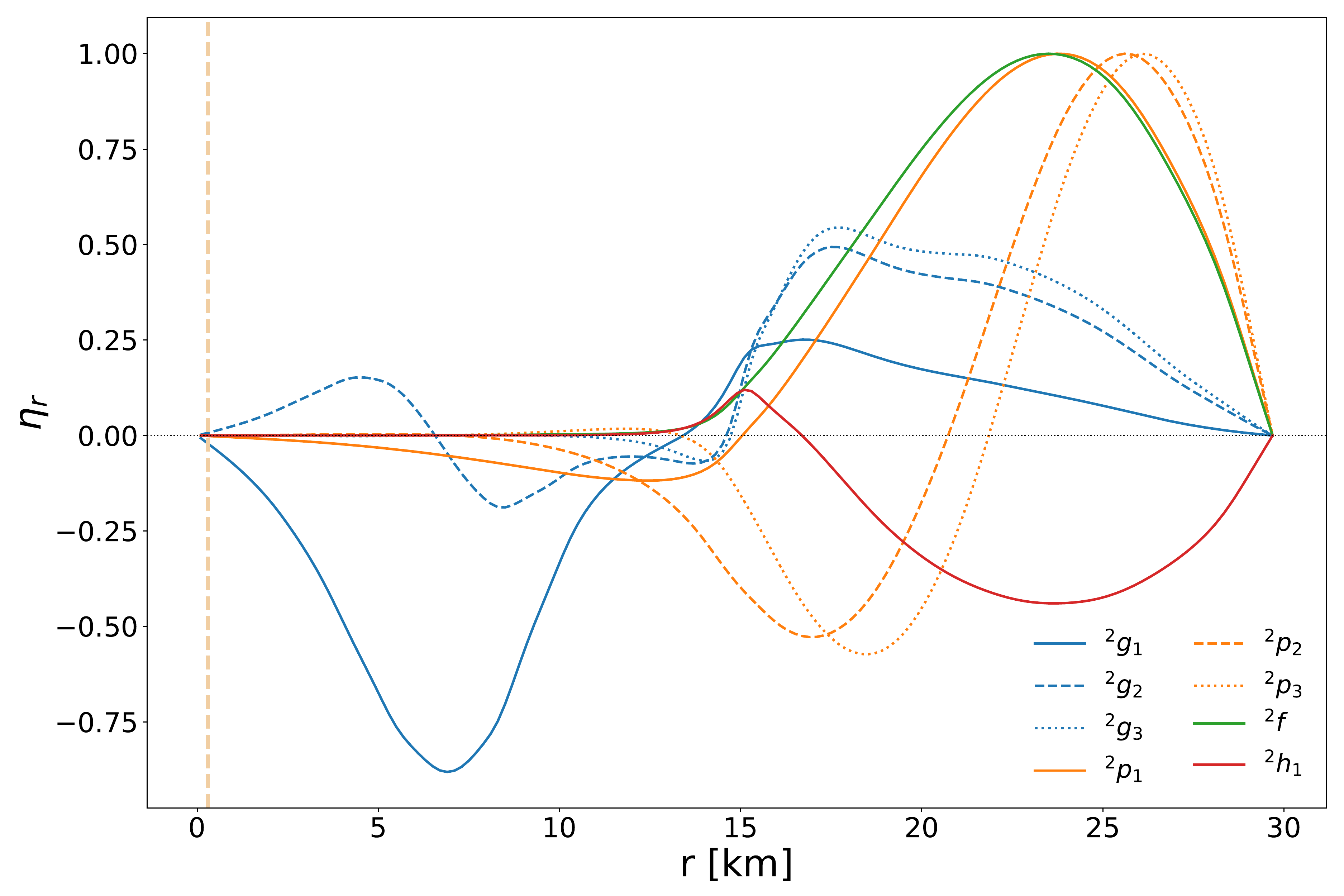}}
		\caption[Radial component ($\eta_r$) of the first modes of each class]{Radial component ($\eta_r$) of the first few modes 
		of each class, $1$~s after bounce. g-modes are represented in blue, p-modes in orange, the f-mode in green, and the 
		h-modes in red. The vertical lines indicate the radial position of the three parts of the star as introduced in Fig.~\ref{fig:regions}. 
		(See main text for details.)  }
		\label{fig:eta_r}
	\end{figure}

\begin{figure}
	\centering
		{\includegraphics[width=.49\textwidth]{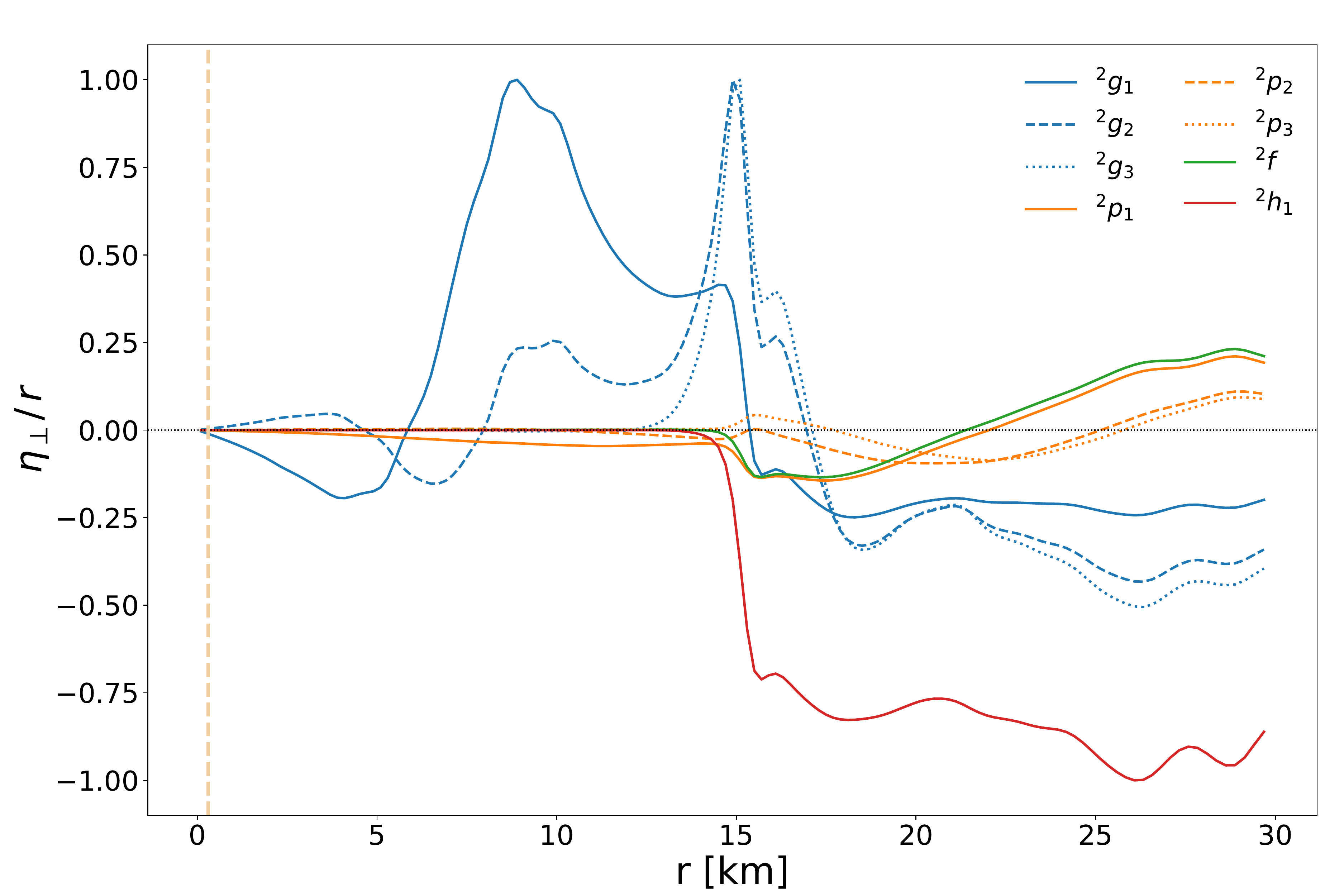}}
		\caption[[Perpendicular component ($\eta_\perp$) of the first modes of each class]{Same as Fig.~\ref{fig:eta_r} but 
		for the perpendicular component ($\eta_\perp$).}
		\label{fig:eta_p}
	\end{figure}

\begin{figure}
	\centering
		{\includegraphics[width=.49\textwidth]{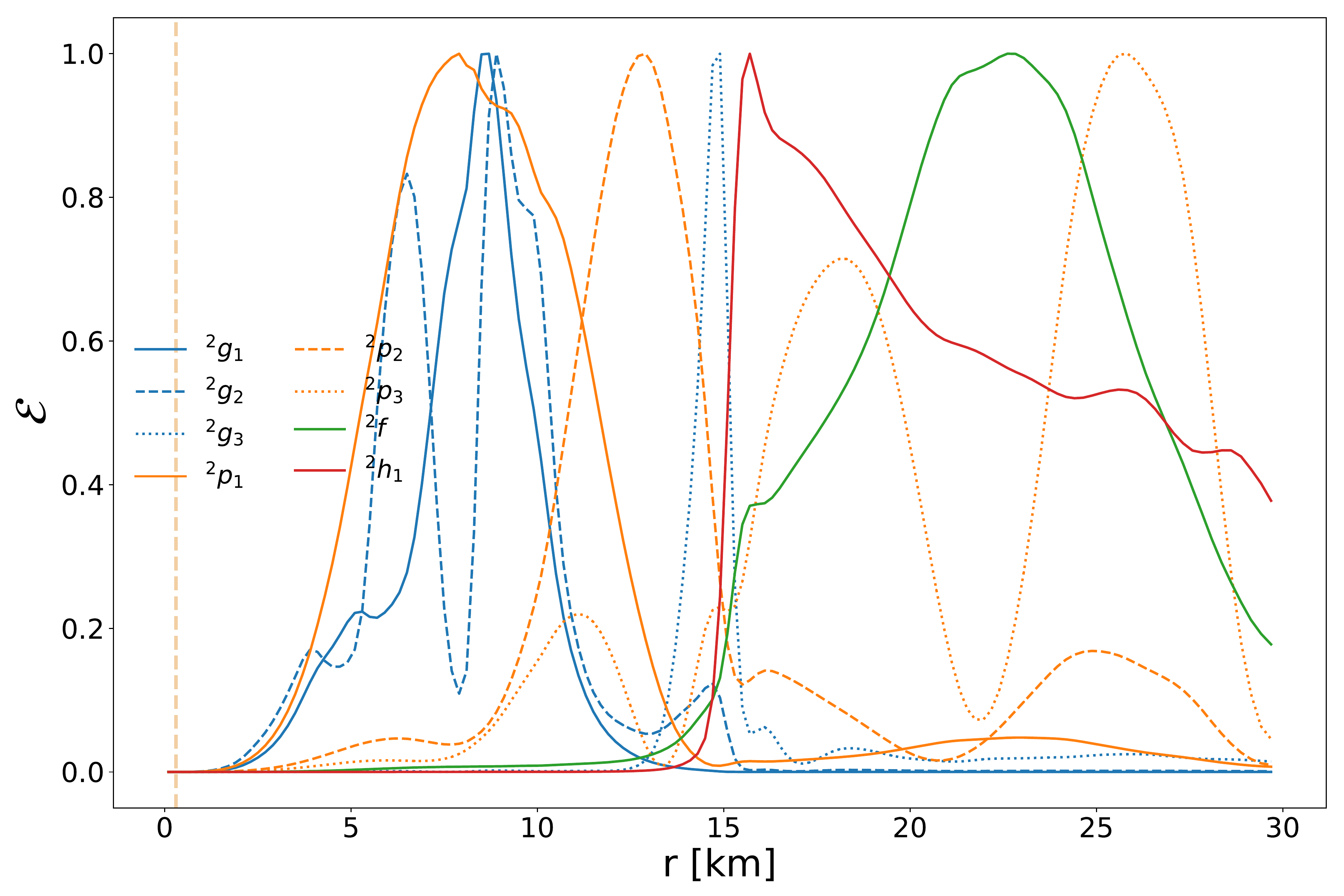}}
		\caption[Energy density $\mathcal{E}$ of the first modes of each class]{Same as Fig.~\ref{fig:eta_r} but 
		for the energy density ($\mathcal{E}$).}
		\label{fig:energy}
	\end{figure}

Once the modes have been identified it is possible to examine in detail their corresponding eigenfunctions. Figs.~\ref{fig:eta_r}, \ref{fig:eta_p} and \ref{fig:energy} show, respectively, the radial component, the perpendicular component, and the energy density of a representative set of the complete modes. Attending to the shape of the modes, one can see that modes within a class are somewhat similar, but with different number of nodes. The vertical dashed lines in Figs.~\ref{fig:eta_r}, \ref{fig:eta_p} and \ref{fig:energy} represent the radial position of the different parts of the star we are considering. The blue line is the limit of the iron core $r_{\rm cold}$. The orange line is the position of the neutrinosphere $r_\nu$ and limits the lower radius of the PNS surface. Finally, the green line represents the radius at which the density is lower than $10^{11}$~g~cm$^{-3}$ and marks the position of the upper radius of the PNS surface. The relative position of each radius at the time shown in these figures is the same as in Fig.~\ref{fig:regions}.

The energy density shown in Fig.~\ref{fig:energy} offers a complementary view to the eigenfunctions, which is useful to 
further interpret the mode classification. All the energy density of the $^2f$ mode is located inside the star, between the 
core and the surface, which we labelled as region II in Fig.~\ref{fig:regions}. The mode \kmt{$^2h_2$} is also confined in the same region, differentiating itself from the g-modes and the p-modes. The g-modes with an odd number of nodes have their energy density confined inside the star, with part of it distributed in regions I and II, while the p-modes with an odd number of nodes have it distributed in region III. The g-mode $^2g_2$ displays a peculiar behaviour as its energy density is confined in region III while the rest of g-modes belong to region I. Also the p-mode $^2p_2$ differs from the others as its energy density is more similar to the h-mode $^2h_2$ than to the other p-modes. It is currently unclear what can be the explanation for the distribution of the energy density among the different regions. Understanding this issue deserves a future study as it could improve our classification procedure.

	\begin{figure}
	\centering
		{\includegraphics[width=.49\textwidth]{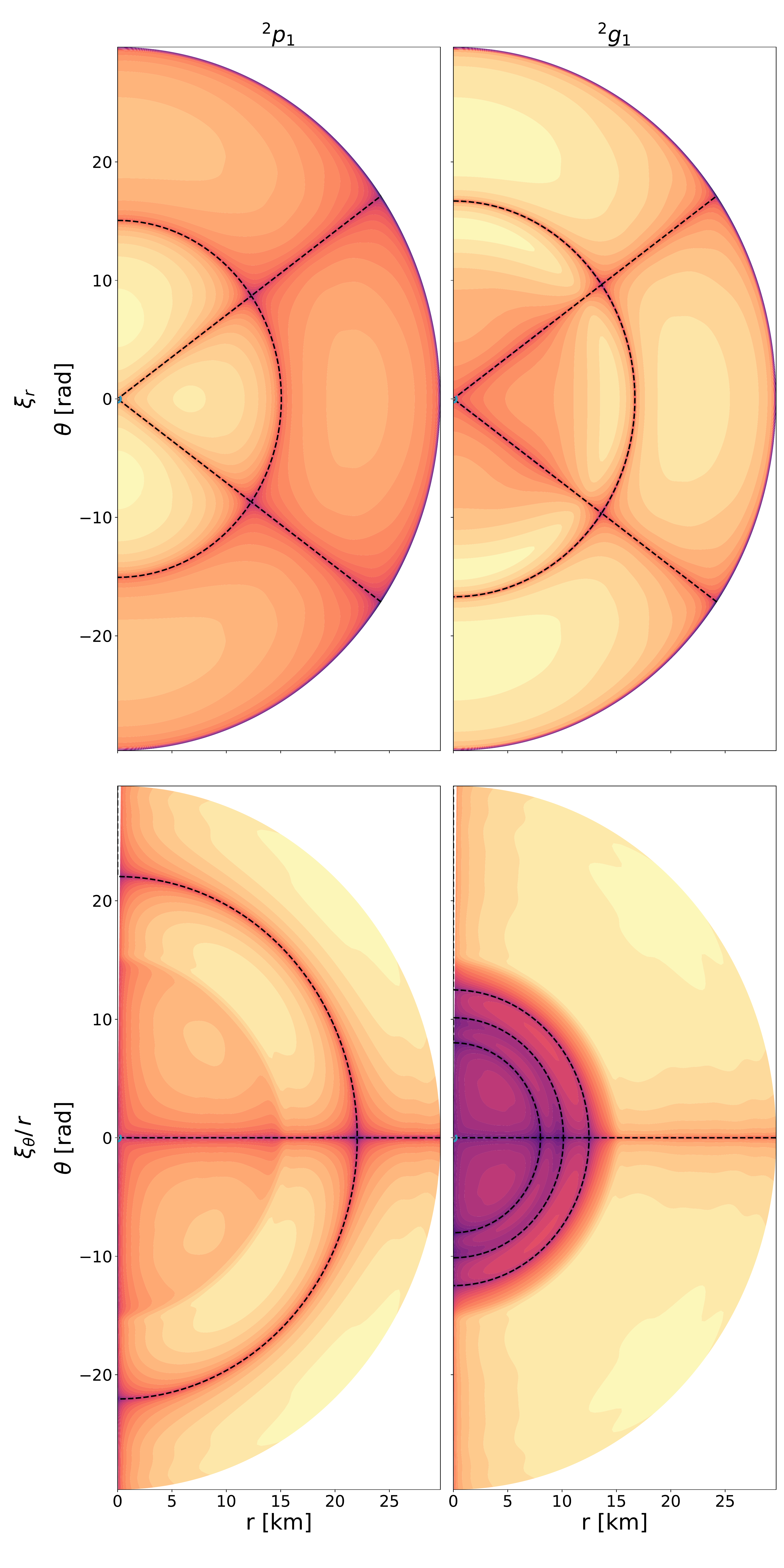}}
		\caption[2D representation of modes $^2p_1$ and $^2g_1$ ]{Polar-coordinates representation of both components of the modes $^2p_1$ and $^2g_1$.}
		\label{fig:2d}
	\end{figure}

To conclude this section we show in Fig.~\ref{fig:2d} the 2D representation of both components of the modes $^2p_1$ and $^2g_1$. The black dashed lines in this figure indicate the zeros of the corresponding eigenfunction. The blue radial line and the orange radial line represent the limits of the PNS core and of the neutrinosphere, respectively. The larger values of the amplitude of each mode are shown in yellow tones, while the lower values are shown in dark red colour. The maximum amplitude of the p-mode (left panels) is concentrated in the exterior part of the PNS, which is consistent with the interpretation of p-modes being standing sound waves trapped between the PNS surface and the shock. However, there is an additional non-negligible component in the interior of the PNS. This component may be due to sound waves penetrating inside the PNS or to the
coupling of the acoustic waves with buoyantly-responding layers in the interior. The right panels show the components of a g-mode that has its maximum at the PNS interior. At this frequency the convectively stable regions in the interior of the PNS allow for the formation of g-modes. However these g-modes extend outside of the PNS up to the shock. The coupling of the PNS surface with the shock is mediated by nodeless sound waves propagating in this region.
	\begin{figure*}
		\centering
		{\includegraphics[width=.95\textwidth]{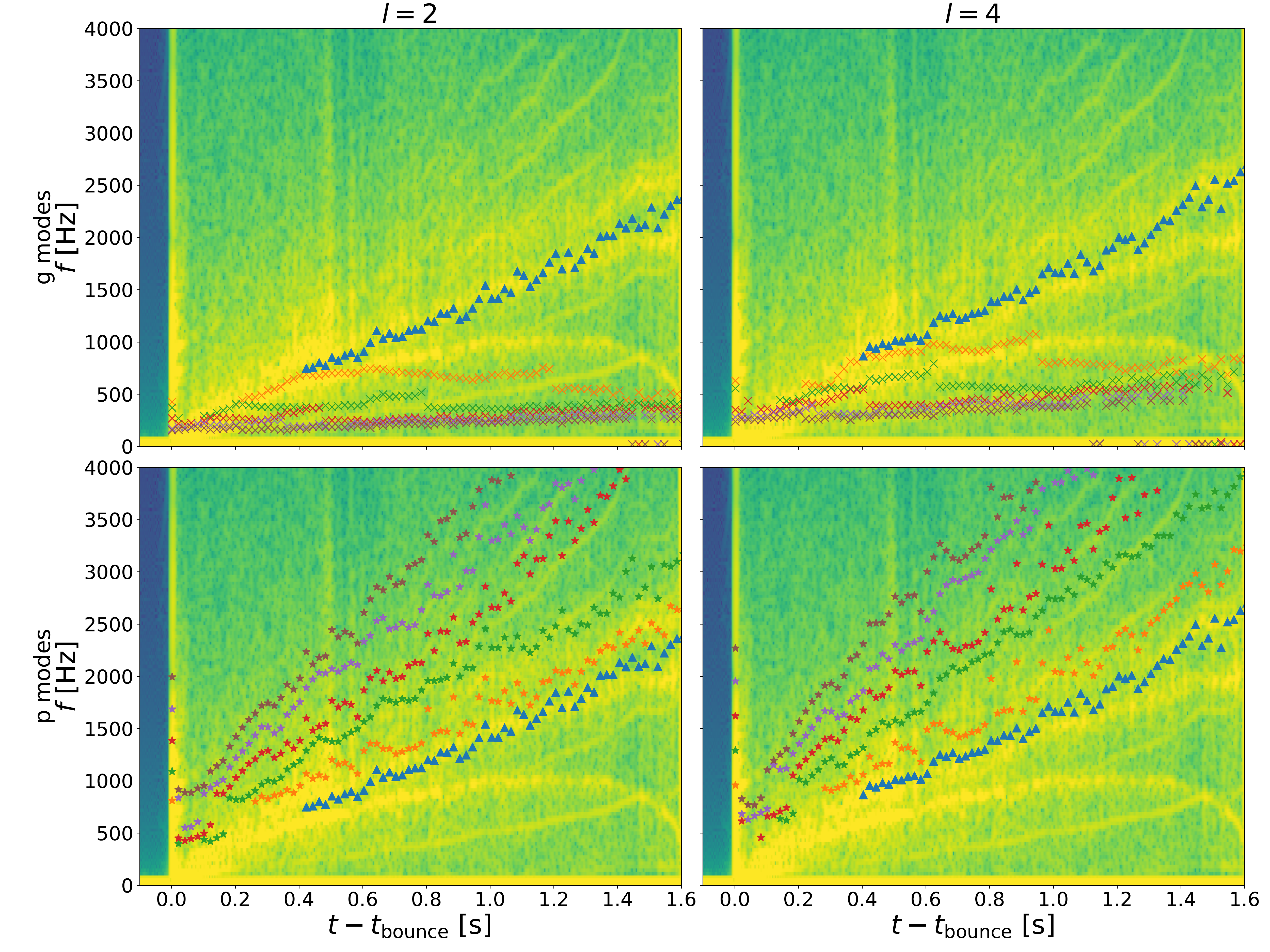}}
		\caption[Comparison of the spectrogram from the numerical simulation with the time-frequency distribution of the $l=2$ and $l=4$ modes.]{Comparison of the spectrogram from the numerical simulation of~\citet{cerda-duran15} with the time-frequency distribution of the $l=2$ and $l=4$ modes. Colour coded is the gravitational-wave power spectral density (PSD), with yellow indicating the highest values. The upper panels show the g-modes (crosses) and the lower panels the p-modes (stars). \kmt{The f-mode (triangles) is shown in both panels}. Left and right panels show the $l=2$ and $l=4$ modes, respectively. The number of nodes are represented in different colours: blue (0), orange (1), green (2), red (3), violet (4), and brown (5).}
		\label{fig:spectrograms}
	\end{figure*}
%
%%%%%%%%%%%%%%%%%%%%%%%%%%%
\section{Gravitational-wave emission}
\label{sec:GW}
%%%%%%%%%%%%%%%%%%%%%%%%%%%
\subsection{Comparison in the frequency domain}
%%%%%%%%%%%%%%%%%%%%%%%%%%%

In the previous sections we have computed the eigenmodes of the system formed by the PNS and the shock at any
given time in the simulation. It is expected that, if we perturb the system, some of those modes will be excited and gravitational waves will be emitted at those particular frequencies. These perturbations are indeed present in the simulation: the region between the PNS surface and the shock is subject to competing instabilities, such as the SASI and convection, that break spherical symmetry. In this region down-flowing plumes of cooled matter hit the PNS surface exciting the eigenmodes of the system. All these perturbations translate into the gravitational-wave signal (see Fig.~\ref{fig:gw_signal}, whose frequency spectrum is expected to be correlated with the frequencies of the modes). In order to compare the two we overplot the time-frequency distribution of the modes with the spectrogram of the gravitational-wave signal obtained from the simulation. For this comparison we consider only modes with $l=2$ and $l=4$. The numerical simulation of~\citet{cerda-duran15} was performed assuming symmetry with respect to the equatorial plane, therefore modes with odd $l$ are not present, and are not considered here. Moreover, $l=0$ modes were not considered in the analysis, and will be considered elsewhere. We do not take into account $l=6$ or higher even $l$ modes because those typically have less energy stored, and hence will produce weaker gravitational-wave signals, than modes with lower $l$.

The results are shown in the spectrograms of Fig.~\ref{fig:spectrograms}. As a general conclusion, most of the features in the spectrograms have a corresponding eigenmode tracing its evolution closely. But the inverse is not true, not every eigenmode computed has a trace in the spectrogram. This is expected, since not every mode has to be excited during the evolution or, even if excited, not every mode can have a sufficiently high amplitude to leave an imprint in the gravitational-wave signal.

%%%%

\kmt{Fig.~\ref{fig:spectrograms} shows that the gravitational-wave emission can be explained with $l=2$ modes alone. Some features may also be explained by the $^4f$ mode (lower-right panel, in blue colour).} During the first half a second after bounce, low order g-modes are the dominant gravitational-wave emission mechanism (upper panels). This is consistent with previous core-collapse simulations, in which spectrogram features during the first half second (previous to the supernova explosion) were interpreted as g-modes \citep[see e.g.][]{Mueller:2013}. At later times, a component of the gravitational-wave signal with increasing frequency appears. This component is perfectly fitted by \kmt{$^2 f$ and} $^2p_n$ modes (\kmt{lower-left} panel) at frequencies larger than $\sim1500$~Hz. $^4p_n$ modes do not appear to contribute to the signal (see discussion in Section~\ref{sec:gweff}).

Below $\sim1500$~Hz, the gravitational-wave emission seems to be dominated by three  features, which coincide in the frequency range
with low order g-modes. However, these features do not have the same shape as g-modes and can be explained in a different manner, as we discuss next. The component whose frequency decreases in time from $\sim 1000$~Hz at about $1.2$~s after bounce to $0$~Hz at the time of BH formation was interpreted in~\citet{cerda-duran15} as a quasi-radial mode ($l=0$). Although $l=0$ modes can be studied in the Cowling approximation (using the appropriate formalism), this approximation do not produce accurate 
results as the object becomes more compact, in particular the frequency of the fundamental $l=0$ mode do not cross zero on the onset of BH formation. Therefore,
we have decided not to consider these modes in this work and leave it for future extensions without the Cowling approximation.
The two features at the lowest frequencies with quasi-linearly increasing frequency (one of them partially hidden by the quasi-radial mode) are discussed in the next section. The apparent lack of g-modes (apart from the fundamental) in the spectrogram is discussed in Section~\ref{sec:gweff}.

In general, it is reassuring that all lowest-order p-modes (in terms of $l$ and $n$) are imprinted in the 
spectrogram. This is consistent with the interpretation that perturbations excite mainly low-order modes and
higher-order modes do not contribute to the signal. Furthermore, this observation also helps to decide which modes are relevant for the gravitational-wave emission, irrespective of the waveforms computed in the simulations. We also note that, at least at high frequencies, the Cowling approximation does not appear to introduce large errors in the eigenmode computation.
Many of the modes show a small mismatch with respect to the spectrogram, typically a small shift to larger frequencies, which is probably caused by the Cowling approximation. The lack of a feature associated with the fundamental mode is puzzling. However, for the case of the fundamental mode the Cowling approximation could introduce large errors, leading to a misinterpretation of the results. In the near future we plan to relax this approximation to analyse whether a closer agreement is found and also to study the behaviour of quasi-radial modes.

%%%%
\subsection{Imprint of the SASI}
%%%%

	\begin{figure*}
	\centering
		{\includegraphics[width=.49\textwidth]{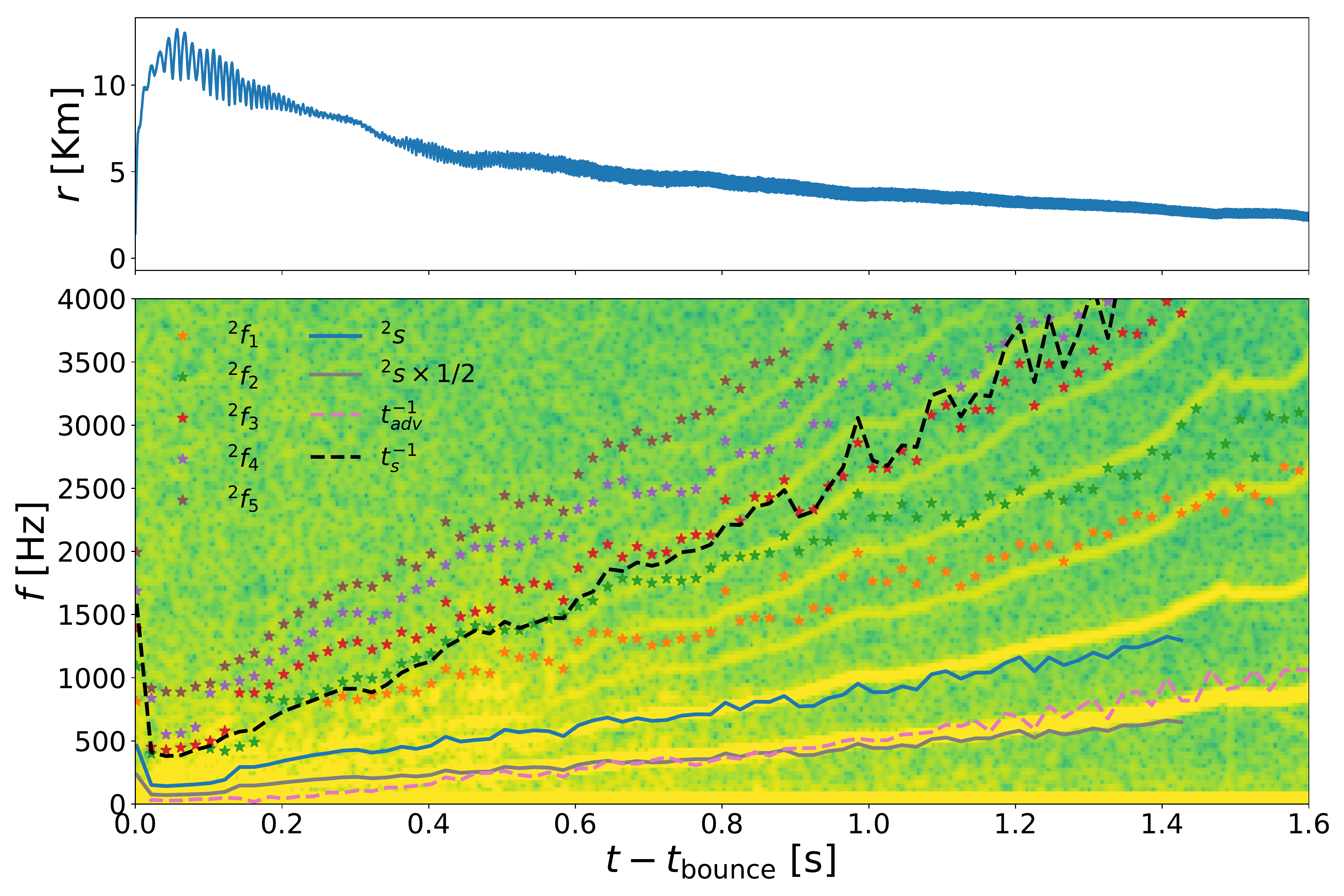}}
		\hspace{-2mm}
	         {\includegraphics[width=.49\textwidth]{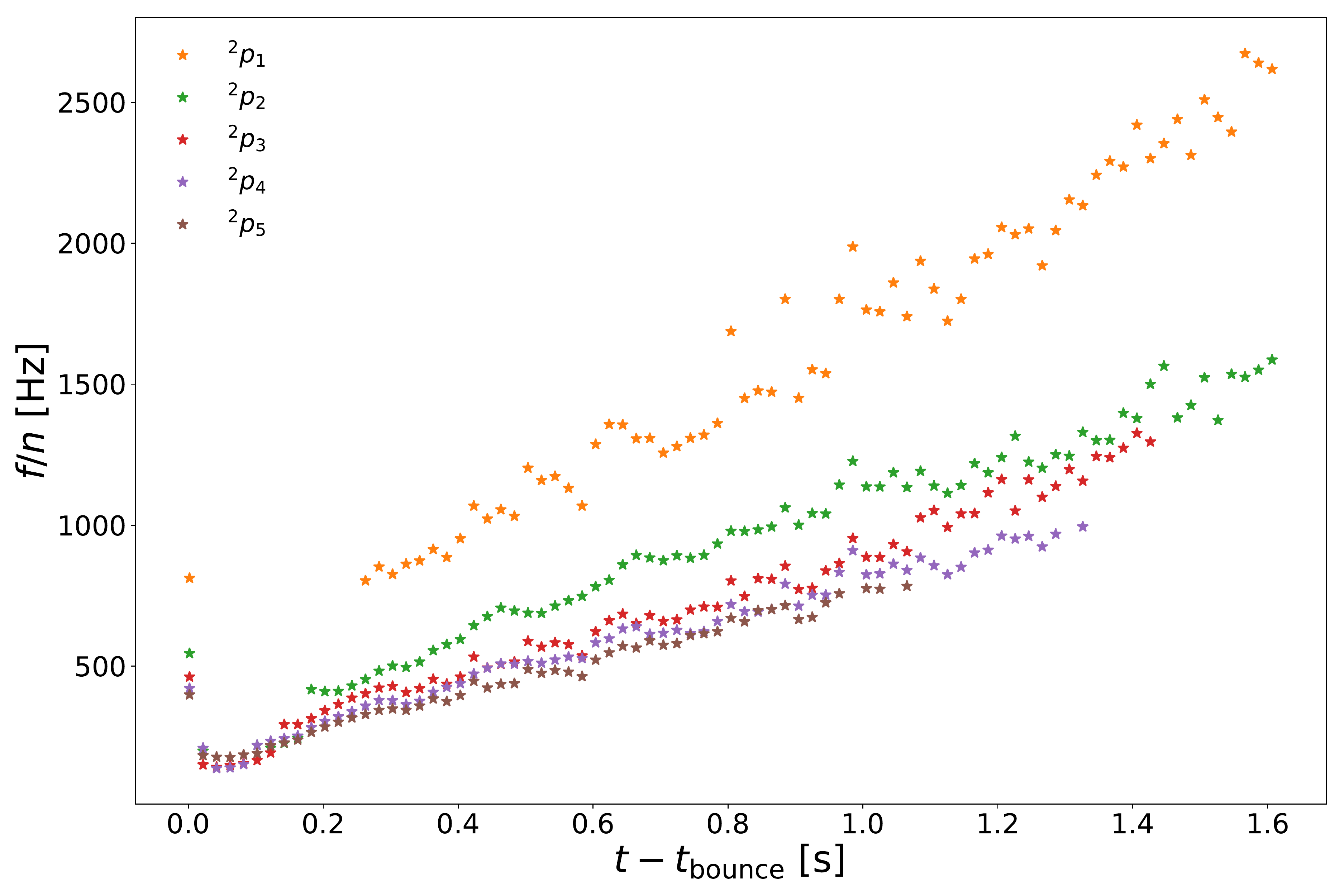}}
		\caption{\kmt{Left panels: time evolution of the radial position of the shock (upper panel) and its corresponding spectrogram 
		(colour coded, lower panel).} Overplotted \kmt{to the latter}, the frequency of the $l=2$ p-modes (symbols)\kmt{, the inverse of the acoustic time, $t^{-1}_{\rm s}$ 
		(dashed black line), 
		the inverse of the advection time $t^{-1}_{\rm adv}$ (pink dashed line)} and integer fractions of the \kmt{characteristic} acoustic frequency: $f_{^2s}$ (solid blue line) and $f_{^2s}/2$(solid magenta line). Right panel: estimation of the $^2s$ frequency using the frequency of $^2p_n$ over $n$. As $n$ increases the frequency converges to that of $^2s$.}
		\label{fig:smodes}
	\end{figure*}
	\begin{figure*}
		\centering
		{\includegraphics[width=.49\textwidth]{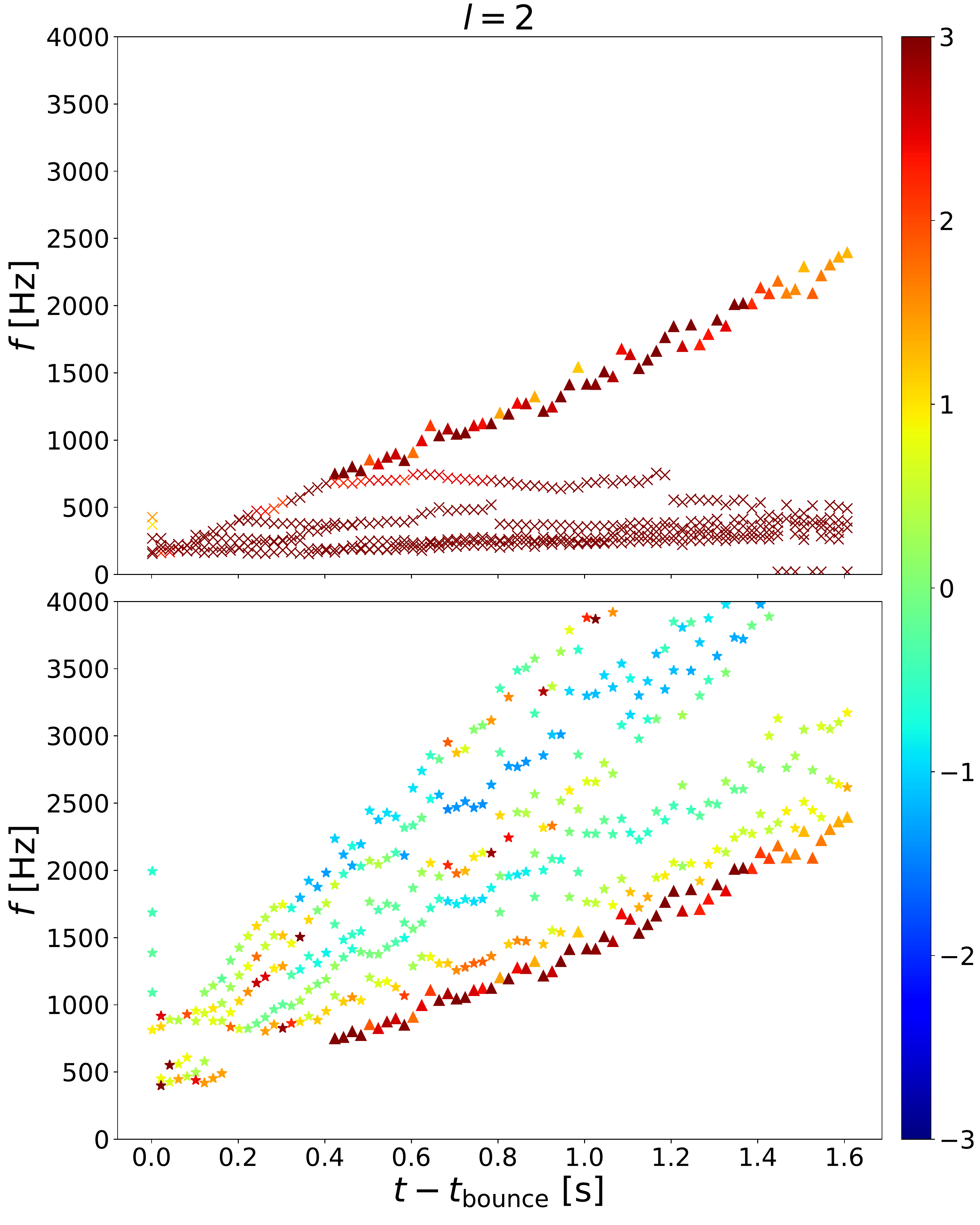}}
		\hspace{-2mm}
		{\includegraphics[width=.49\textwidth]{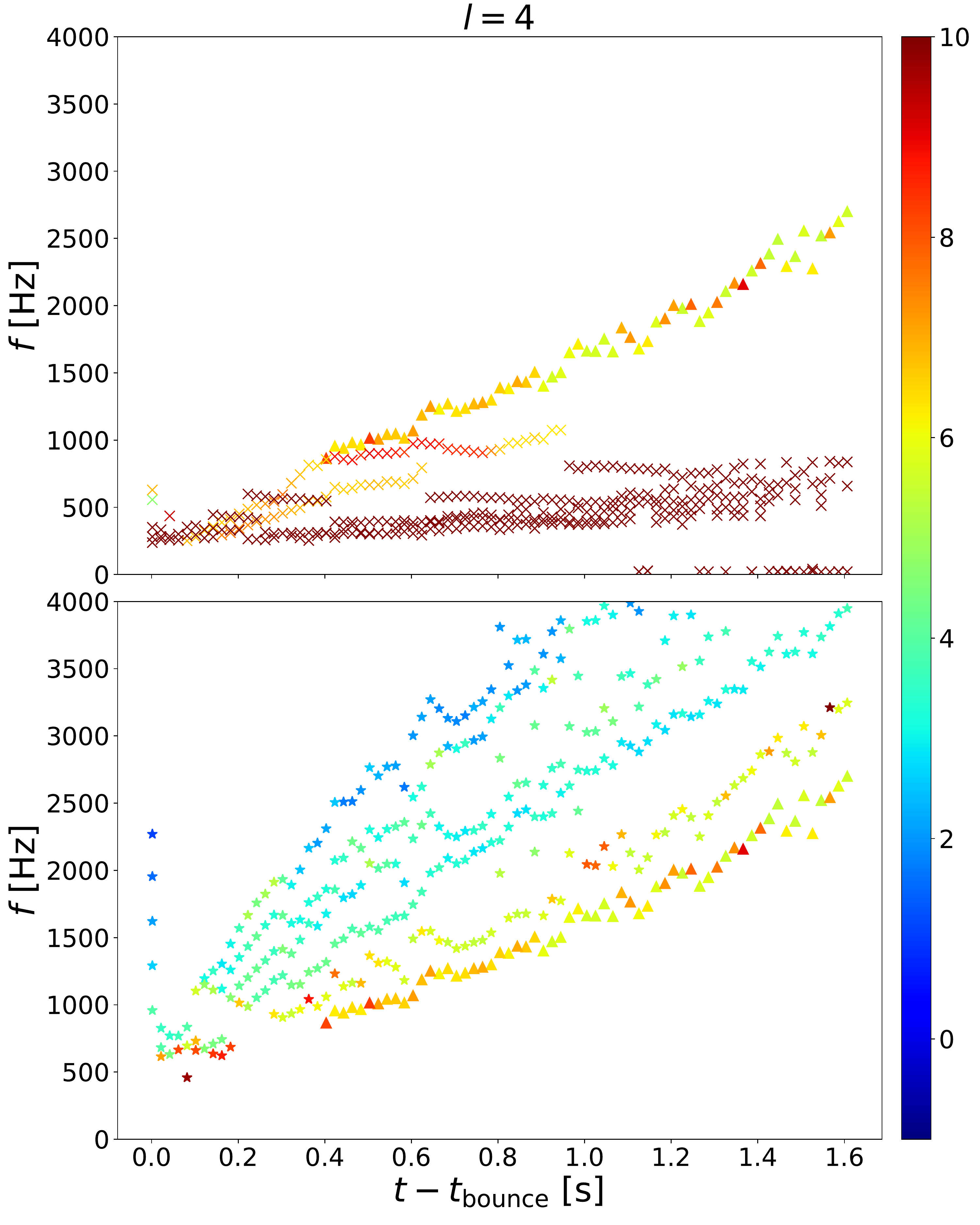}}
		\caption[Time-frequency diagram of all (complete) modes separated by $\tau_{\rm GW} $.]{Time-frequency diagram of the
		 modes of the complete problem separated by $\tau_{\rm GW}$. Modes with lower values of $\tau_{\rm GW} $ are coloured in blue and
		 the values scale to the larger values in red. The upper row corresponds to g-modes (crosses), while the lower row corresponds to p-modes (stars). The
		 fundamental mode is represented with triangles. The colour bar shows the logarithm of $\tau_{\rm GW}$, in seconds.}
		\label{fig:tf-tau}
	\end{figure*}

\kmt{Most of the} features \kmt{identified as} p-modes in this work were identified as SASI modes by \citet{cerda-duran15}  because their frequencies trace the sloshing motions of the shock. This can be seen in the \kmt{lower} left panel of Fig.~\ref{fig:smodes}, where we plot a spectrogram of the radial position of the shock at the equator \kmt{and compare with the p-mode frequencies (symbols).} It is interesting to observe that, for frequencies below the lowest order p-mode, there are features in the spectrogram, both of the gravitational-wave signal and of the shock location, similar in shape to the p-mode signatures but at lower frequencies. These low-frequency signals were attributed to SASI in \citet{cerda-duran15} as well as in~\citet{Kuroda:2016} and~\citet{Andresen:2017}. The fact that they appear in the spectrogram of the shock location (while the fundamental frequency does not) confirms that they are not likely associated with g-modes, despite the similar frequencies, but with something else.

To understand these features we briefly review the mechanism responsible for the SASI \citep{Foglizzo:2000,Blondin:2003}. Given a perturbation at the location of the shock, this is advected downwards by the subsonically accreting fluid until it reaches the surface of the PNS, at which point it decelerates and can excite sound waves that travel upwards, further exciting the shock. Under the right conditions \citep[see][]{Foglizzo:2002,Foglizzo:2007}, this advective-acoustic cycle is unstable and the amplitude of small perturbations grows exponentially into the non-linear regime. In the same region, between the PNS and the shock, it is also possible to form a purely acoustic cycle, in which both the downward and upward perturbations are acoustic. This acoustic cycle has been shown to be stable under the conditions present in this scenario~\citep{Foglizzo:2007}. 

\kmt{Our linear analysis allows us to study the purely acoustic cycle, which corresponds to the computed p-modes. However, we cannot obtain SASI frequencies 
in our framework because we are neglecting the sub-sonic accreting flow below the shock (we consider hydrostatic equilibrium). In any case, it is clear that our analysis shows that the SASI present in the simulation is not 
caused by a purely acoustic cycle (with higher frequencies associated to the p-modes) and hence it is likely produced by an advective-acoustic cycle.} 

\kmt{To interpret the modes of the lower left panel of Fig.~\ref{fig:smodes} in terms of advective-acoustic and purely acoustic cycles we plot the inverse of the acoustic and advective
times 
\begin{equation}
t_{\rm s} = \int_{r_{\rm PNS}}^{r_{\rm shock}} \frac{dr}{\alpha c_s} \quad ; \quad  t_{\rm adv} = \int_{r_{\rm PNS}}^{r_{\rm shock}} \frac{dr}{\alpha |v_r|},
\end{equation}
respectively (dashed lines on top of the spectrogram). Since $t_{\rm s}$ is the sound crossing time of the region between the PNS and the shock, the value of $t_{\rm s}^{-1}$
matches the typical frequencies and rising behaviour of all p-modes. However, since it is not well determined the exact location at which sound waves are reflected
at the PNS surface, we cannot use this value reliably as an estimator for the p-mode frequency. Alternatively, one can estimate the frequency of the purely-acoustic
cycle directly from the p-mode frequencies. For sufficiently large $n$ (and small $l$), we are close to the short-wavelenght (geometric) approximation of waves and p-modes are close 
to an integer number of times a characteristic acoustic frequency $f_{^2s}$.
On the right panel of Fig.~\ref{fig:smodes} we estimate this frequency by plotting $^2p_n$ modes divided by $n$. We use $n=3$ to estimate the value of $f_{^2s}$ in our system.}

\kmt{The two features in the spectrogram  (lower left panel of Fig.~\ref{fig:smodes}) not explained by p-modes can be explained in terms of the advective-acoustic cycle.
Given that the acoustic time is much smaller than the advective time, the frequency of the advective-acoustic cycle is approximately $t_{\rm adv}^{-1}$. This frequency 
matches nicely the lowest observed frequency. The second feature has a frequency of $\sim 2t_{\rm adv}^{-1}$ and is possibly an overtone.
Additionally, we observe that these two low-frequency SASI features correspond to $\sim f_{^2s}$ and $\sim f_{^2s}/2$. This could be an indication of a constructive
interference between the acoustic-advective and the purely acoustic cycles as proposed by \cite{Fernandez:2009}, however a more detailed analysis of the system should
be done to confirm this fact.}

%%%%%%%%%%%%%%%%%%%%%%%%%%%%%
\subsection{Gravitational-wave radiation efficiency}
\label{sec:gweff}
%%%%%%%%%%%%%%%%%%%%%%%%%%%%%
%
	\begin{figure*}
	\centering
		{\includegraphics[width=.49\textwidth]{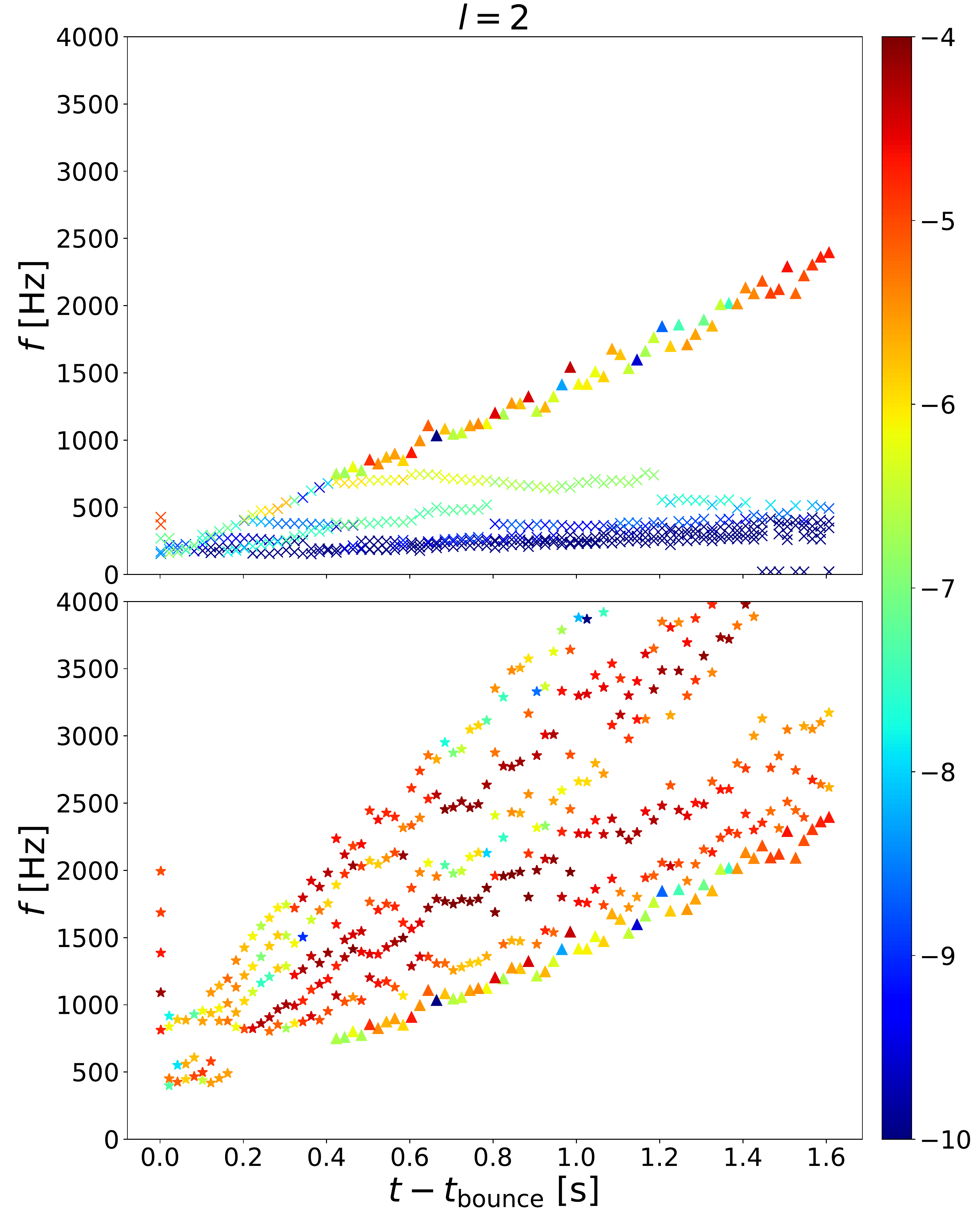}}
		\hspace{-2mm}
	         {\includegraphics[width=.49\textwidth]{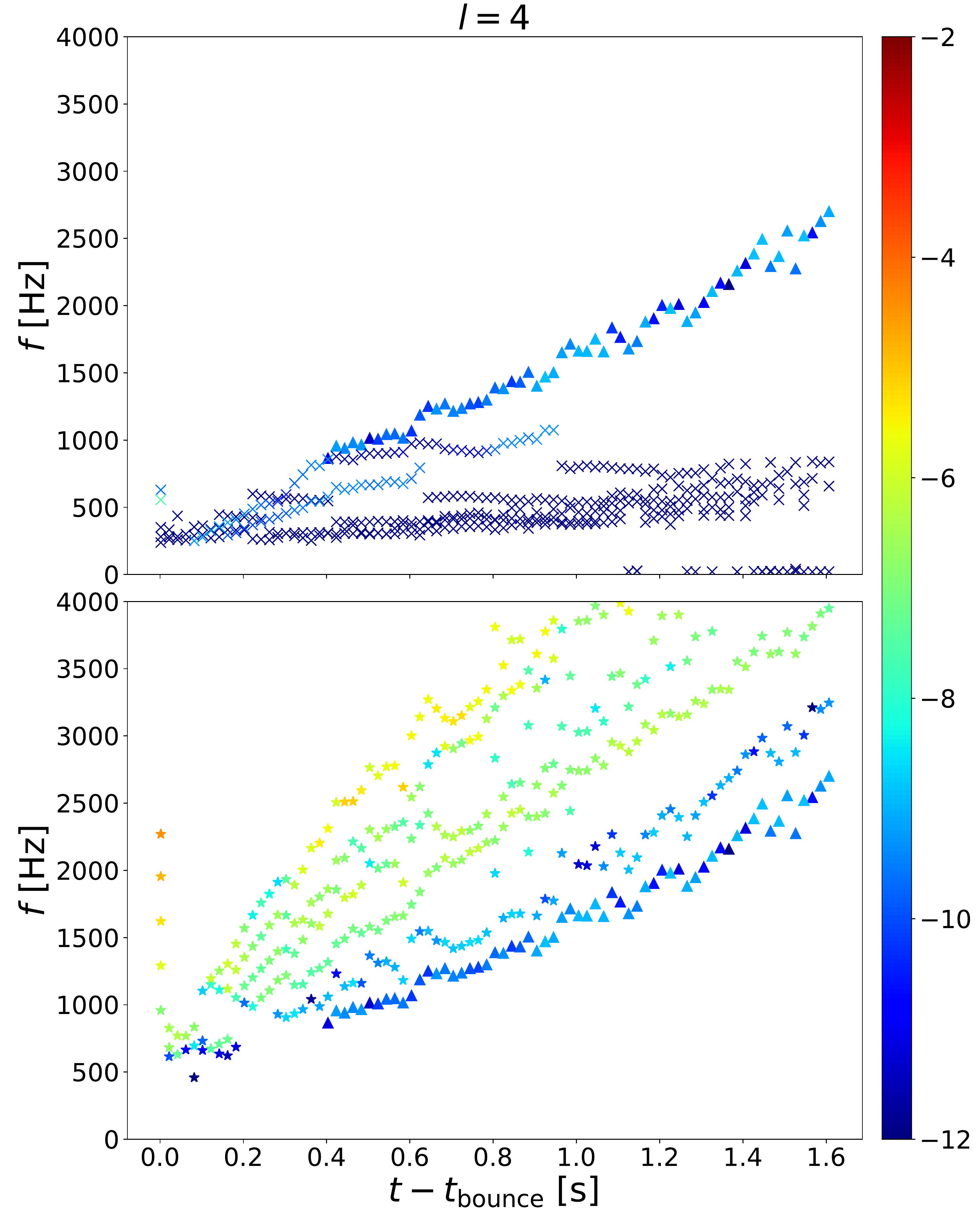}}

		\caption[Time-frequency diagram of all (complete) modes separated by their efficiency.]{Time-frequency diagram of the
		 modes of the complete problem separated their efficiency. Modes with low values are coloured in blue and
		 their values scale to the larger values in red. The upper row corresponds to g-modes (crosses), while the lower row 				corresponds to p-modes (stars). The fundamental mode is represented with triangles. The colour bar shows the logarithm 		of the efficiency.}
		\label{fig:tf-efficiency}
	\end{figure*}
In the previous sections we have compared our eigenmode analysis with the gravitational-wave signal only in terms of frequency evolution. Unfortunately, the eigenmode analysis does not allow us to predict what would be the amplitude of each mode in
the gravitational-wave signature. This amplitude depends on two factors: the energy stored in each mode and the efficiency of each mode to emit gravitational waves. In the previous section we have also found indications that most of the energy of the perturbations is stored in the low-order modes. The energy $E$ of a mode can be computed from Eq.~(\ref{eq:energy2}) and the radiated power $P$ is given by Eq.~(\ref{eq:power}). Both quantities are proportional to the (unknown) eigenmode amplitude. The ratio between both quantities
\begin{equation}
\tau_{\rm GW} = \frac{E}{P},
\end{equation}
is the gravitational-wave timescale, which gives an idea of the characteristic time in which the energy of a 
mode will be radiated in gravitational waves. Modes with smaller value of $\tau_{\rm GW}$ are expected to be more efficient 
emitters of gravitational waves and will produce a stronger imprint in the signal. We can also define the gravitational-wave emission 
efficiency as
\begin{equation}
\textrm{(GW efficiency)}= \frac{P}{E f},
\end{equation}
where $f$ is the frequency of the mode. This equation gives an idea of the fraction of the mode energy radiated in gravitational waves per oscillation cycle.

Fig.~\ref{fig:tf-tau} shows the time-frequency diagram of $\tau_{\rm GW}$ for each mode. 
Modes with the lowest values of $\tau_{\rm GW}$ (colour coded) are the strongest, and most efficient, emitters of gravitational waves (see Fig.~\ref{fig:tf-efficiency}).
Comparing this figure with Fig.~\ref{fig:spectrograms}, one can see that the $^2p_4$ mode (light-blue stars in Fig.~\ref{fig:tf-tau})
is the most efficient emitter. 
From the definition of $\tau_{\rm GW}$, one realises that 
\begin{equation}
\tau_{\rm GW} \propto \frac{(n (2l+1)!!)^2}{\sigma^{2l}},
\end{equation}
where the dependence with $n$ is a crude approximation and a consequence of the different number of nodes in the integrands 
of Eqs.~(\ref{eq:energy2})  and (\ref{eq:power}). 
This dependence produces a qualitatively different behaviour of p-modes and g-modes in terms of $\tau_{\rm GW}$.
In both cases, the gravitational-wave emission of high-order $n$ modes is expected to be suppressed. However,
for p-modes, $\tau_{\rm GW}$ decreases with increasing frequency. This compensates somewhat the suppression factor 
at high $n$ and allows the appearance of high order p-modes up to $n=5$ in the spectrogram. On the contrary, 
gravitational-wave emission in high-order g-modes is suppressed by both the high value of $n$ and the low frequency, which results
in no observed g-mode frequencies apart from the fundamental. It seems clear that our classification procedure allows us to match the features in the spectrogram with modes with low values of $\tau_{\rm GW}$, i.e. with high efficiency.

Regarding $l=4$ modes, we find that $^4p_4$ is the mode with the lowest $\tau_{\rm GW}$. 
Note that when comparing $l=2$ and $l=4$ modes, there is a difference in $\tau_{\rm GW}$ of several orders of magnitude.
The reason for this difference is that, for perturbations of a spherical background, $l=4$ modes do not contribute to 
the quadrupolar contribution of the gravitational-wave emission, but exclusively to the octupolar component. This component  is highly suppressed by a $1/c^2$ factor with respect to $l=2$ modes\footnote{{This factor cannot be seen explicitly in Eq.~(\ref{eq:power}) because of the use of units with $c=1$.}}. However, the simulation of~\citet{cerda-duran15} showed that the PNS is deformed by rotation and hence the $l=4$ mode is able to contribute to the quadrupolar radiation. This enhances emission from these modes, which may explain some of the features in the spectrogram. Therefore, in order to properly capture the gravitational-wave emission properties of modes with $l\ne 2$ in rotating cores, one should  take into account corrections due to to deformations of the PNS. These corrections are out of the scope of the present qualitative analysis and will be investigated in the future. We can also conclude that, for nonrotating cores, in which the background is spherically symmetric, modes with $l\ne2$ cannot contribute significantly to the gravitational-wave emission, and one could restrict the analysis to $l=2$ modes. 

In spite of the fact that the $^2p_4$ mode is the most efficient emitter, the most prominent feature in the spectrogram is actually closer to the $^2p_1$ mode.  This means that gravitational-wave efficiently alone cannot explain the relative amplitude of the different features in the spectrogram, but one has also to consider the relative amplitude on the energy stored in the different modes excited in the system. If one considers the relative amplitude of different features in the PSD of the radial location of the shock (left panel of Fig.~\ref{fig:smodes}) as a measure of the relative difference of the energy in the different modes, one can see that for higher order modes (higher $n$) the amplitude is smaller. The decay with $n$ scales roughly as $(n+1)^{-5}$, which is steeper than that of a 2D turbulent cascade which goes as $(n+1)^{-3}$. This steep decline could explain why the most prominent feature in the gravitational-wave spectrogram is associated with the $^2p_1$ mode instead of the most efficient mode ($^2p_4$). As noted by \cite{Andresen:2017}, the distribution of energy among different components of the gravitational-wave signal can depend strongly on the dimensionality of the simulation (2D or 3D). Therefore, in order to be able to predict which is the most dominant mode observed in the gravitational-wave signal, it will be necessary to perform more detailed 3D simulations of the scenario to understand how energy is distributed among different modes. 

%%%%%%%%%%
\section{Summary}
\label{sec:summary}
%%%%%%%%%%

In this paper we have taken first steps towards the development of asteroseismology in the core-collapse supernova scenario using gravitational-wave observations. The ultimate goal of this investigation is the possibility of inferring astrophysical parameters of the resulting proto-neutron stars. Our approach, based on linear-perturbation analysis in full general relativity, has been applied to compute the eigenmodes of oscillation of a background model which has been taken from the axisymmetric core-collapse simulation of~\citet{cerda-duran15}. The physical system defining the cavity where modes are computed is defined by the region extending from the centre of the PNS to the shock location, and includes the convectively unstable hot-bubble region. 

We have explored if the spectrum of the gravitational-wave signal produced as a result of the collapse of the core of a 35$M_{\odot}$ star, from the bounce signal to the $\sim 2$ s long post-bounce signal, associated with an accretion phase leading to the formation of a black hole, can be related to the modes of vibration of the system. To do so we have computed the number of nodes of each of the oscillation modes found, either (buoyancy-mediated) g-modes, (pressure-mediated) p-modes, and hybrid modes (a mixture of the two). Our simplistic mode classification in terms of the number of nodes has shown the presence of those sets of modes. We have found that for the p-modes the  number of nodes increases with frequency while g-modes follow the opposite trend. 
The two families are neatly separated by the so-called fundamental mode, a mode with no radial nodes. The comparison of our perturbative results with the fully nonlinear results of the numerical simulation has allowed for an unambiguous identification of the two groups of modes. Further support has been collected through the use of the spectrogram of the gravitational-wave signal obtained in the numerical simulation, which has revealed a remarkable correspondence between both results. 

We have also confirmed  the result obtained by previous works~\citep{cerda-duran15,Kuroda:2016,Andresen:2017}, showing that some low-frequency features are associated with the SASI and correspond to \kmt{the inverse of the advection timescale in } the region between the PNS surface and the shock
These SASI modes are of particular interest because their frequencies, specially in the first half-second of the evolution, lay close ($\sim 100$~Hz) to the maximum sensitivity of ground-based interferometers as Advanced LIGO, Advanced Virgo, and KAGRA.

The analysis presented in this paper provides a proof-of-concept that asteroseismology is indeed possible in the core-collapse scenario, despite the complexities of the system, and it will serve as a basis for future work on PNS parameter inference based on gravitational-wave observations. The ultimate goal of our study is to develop a reliable model to relate the parameters of the core-collapse progenitor with the corresponding spectrum of oscillation of the PNS, and therefore with the gravitational-wave spectrum, that allows us to infer those parameters directly from the gravitational-wave spectrum without the need to perform computationally expensive multidimensional simulations. Our immediate plans involve the simulation of a large set of (cheap) one-dimensional core-collapse models spanning the parameter space of the progenitors in order to study the dependence of the time-frequency distribution and power of the modes with those parameters. Furthermore,  the assumption of the Cowling approximation taken in the current work will be relaxed, and our method will be also assessed with existing results from three-dimensional core-collapse simulations.

\section*{Acknowledgements}
Work supported by the Spanish MINECO (grant AYA2015-66899-C2-1-P) and the Generalitat Valenciana (PROMETEOII-2014-069). 
A.P.~acknowledges support from the European Union under the Marie Sklodowska Curie Actions Individual Fellowship, grant agreement no 656370.
\kmt{We thank T. Foglizzo for useful discussions.}

%%%%%%%%%%%%%%%%%%%%%%%%%%%%%%%%%%%%%%%%%%%%%%%%%%

%%%%%%%%%%%%%%%%% APPENDICES %%%%%%%%%%%%%%%%%%%%%

\appendix

\section{Oscillation modes of a constant-density sphere}
\label{sec:appendix}

As a test for our eigenvalue solver we consider a sphere of radius $R=1$, constant density (we set $\rho=1$ for simplicity) and constant speed of sound, $c_{\rm s} =1$. The solution for this problem can be obtained analytically and reads,
\begin{align}
\label{eq:analitica}
	\eta_r &= \eta_0 \, \partial_r (j_l (\sigma_n r))\,, \\
	\eta_\perp &= \eta_0 \, j_l (\sigma_n r)\,,
\end{align}
where $j_l$ is the spherical Bessel function of the first kind and $\eta_0$ is a constant describing the amplitude of the mode. 
Moreover, $\sigma_n$ are the eigenvalues, which correspond to those values that make $\eta_r(r=1)=0$, i.e. the roots of $j_l(\sigma_n)$.

\begin{figure}
	\centering
		{\includegraphics[width=.49\textwidth]{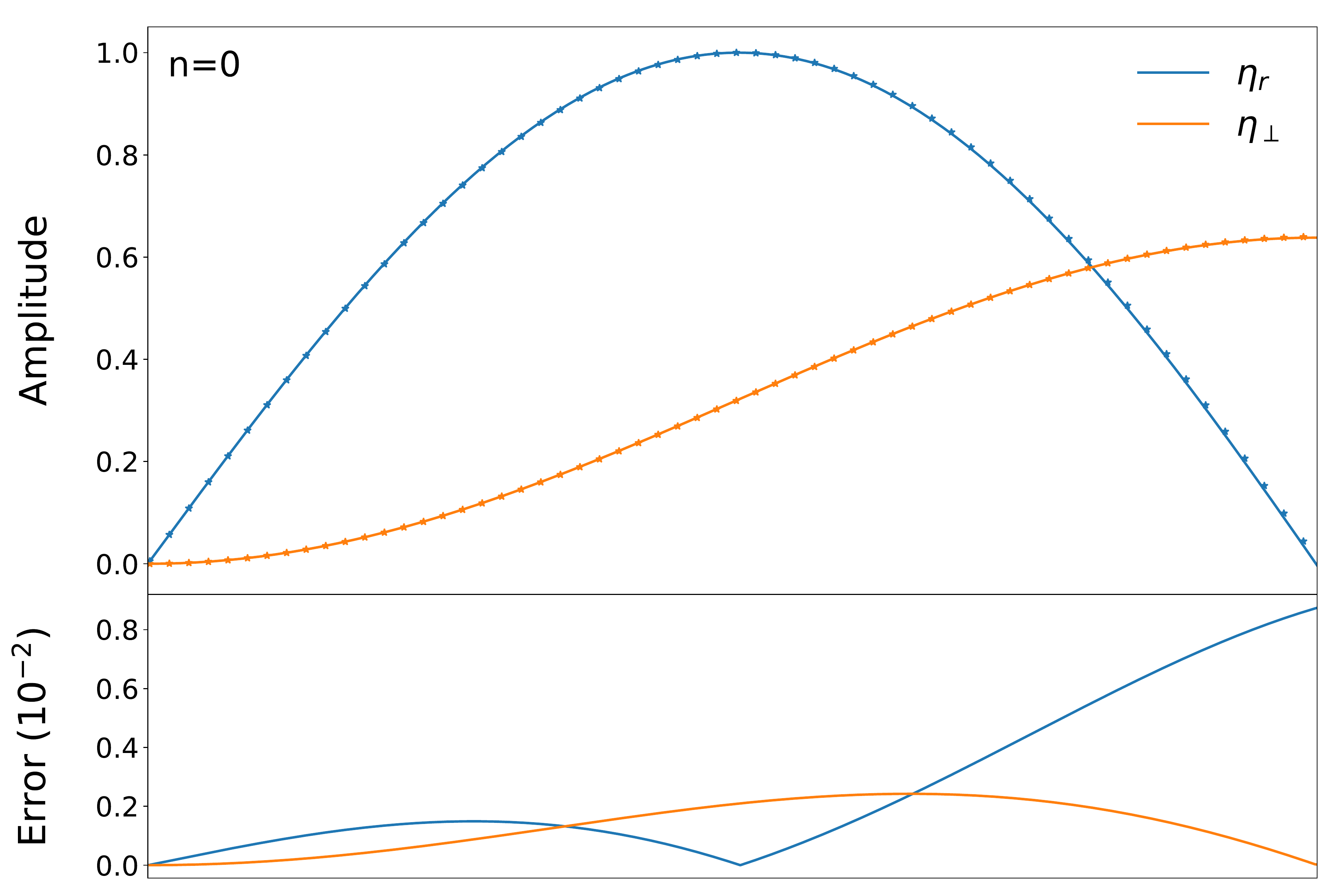}}\\
		{\includegraphics[width=.49\textwidth]{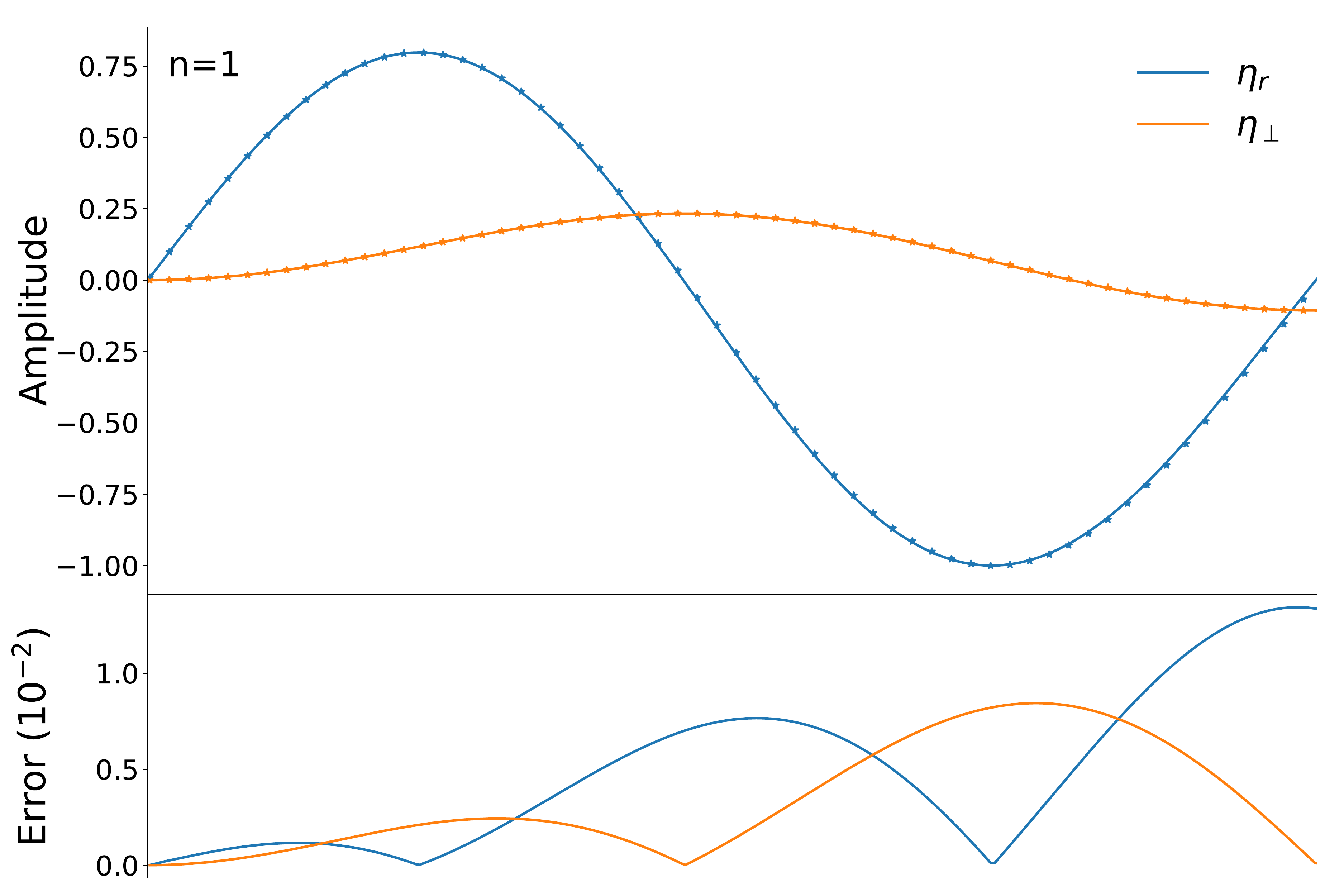}}\\
		{\includegraphics[width=.49\textwidth]{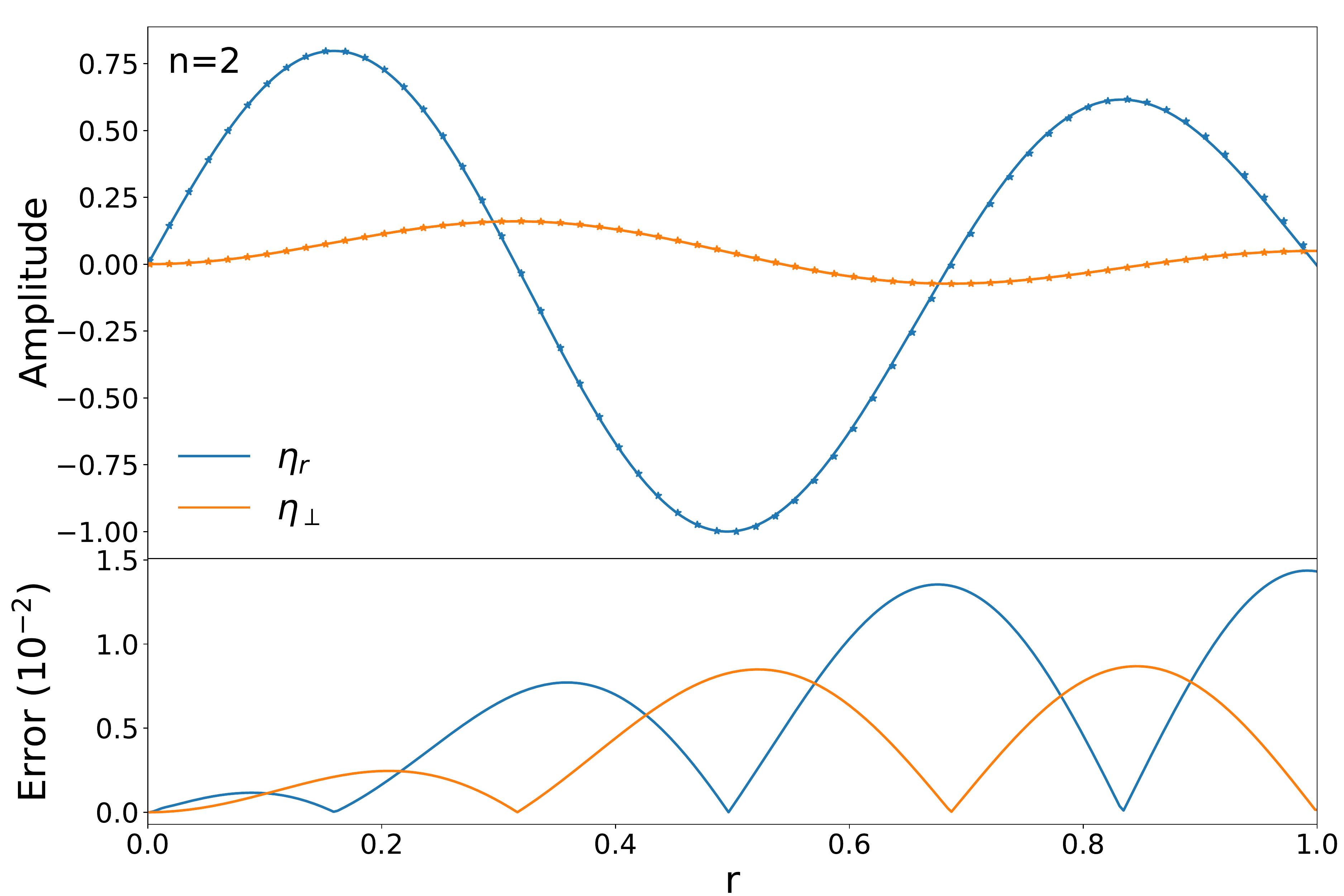}}
		\caption[Caption.]{Comparison between the analytic eigenfunctions (solid lines) and the numerical eigenfunctions (symbols) for the first three modes of oscillation of a constant-density sphere. The lower subpanels of each plot show the relative errors.}
		\label{fig:analytic}
	\end{figure}

We have computed the eigenvalues and eigenfunctions using our numerical method and have compared them with the analytical solution. For this purpose we use an equally-spaced grid with $N=300$ zones in the interval $[0,1]$, which is comparable to the number of points inside the shock in the numerical simulation of the model 35OC of~\citet{cerda-duran15}. Fig.~\ref{fig:analytic} shows the first three eigenfunctions, computed both numerically and analytically, as well as the corresponding relative errors. The numerically computed eigenvalues for the first seven modes are reported in Table \ref{tab:comp}. In all cases the root-mean-square error (RMSE) in the computation of the frequencies, defined as $\sqrt{\frac{1}{N}\sum_{i=0}^N (f_{\rm a} - f_{\rm n})^2}$, where subindices 'a' and 'n' refer to the analytic and numerical frequencies, respectively, is at most of ${\mathcal O} (10^{-3})$.  

\begin{table}
\label{tab:comp}
\caption{Comparison between the analytic and numerical oscillation frequencies (first seven modes) of a constant-density sphere. The columns report the analytic frequency, $f_{\rm a}$, the numerical frequency, $f_{\rm n}$, their relative error, and the RMSE for radial solutions and perpendicular solutions for all first seven modes.}
\begin{center}
\begin{tabular}{ccccc}
$f_{\rm a}$ & $f_{\rm n}$ & $\displaystyle{\frac{f_{\rm a}-f_{\rm n}}{f_{\rm a}}}$ & $\rm{RMSE}_r$ & $\rm{RMSE}_\perp$ \\
& & & $(10^{-3})$ & $(10^{-3})$\\
\hline
3.34 & 3.35 & 0.00098& 2.35 & 0.33 \\
7.29 & 7.30 & 0.001& 3.93 & 0.70 \\
10.61 & 10.62 & 0.001& 4.70 & 0.53 \\
13.85 & 13.86 & 0.0011& 5.05 & 0.42 \\
17.04 & 17.06 & 0.0012& 5.28 & 0.34 \\
20.22 & 20.25 & 0.0012& 5.43 & 0.29

\end{tabular}
\end{center}

\end{table}



\bibliographystyle{mnras}
\bibliography{./references} % if your bibtex file is called example.bib

%\begin{thebibliography}{99}
%
%\bibitem[Isenberg(2008)]{Isenberg08} Isenberg, J.~A.\ 2008, International Journal of Modern Physics D, 17, 265 
%
%\bibitem[Wilson et al.(1996)]{Wilson96} Wilson, J.~R., Mathews, G.~J., \& Marronetti, P.\ 1996, \prd, 54, 1317 
%
%
%\bibitem[Abdikamalov(2014)]{abdikamalov14}{Abdikamalov}, E. and {Gossan}, S. and {DeMaio}, A.~M. and {Ott}, C.~D.
%\bibitem[Janka(2007)]{Janka:2007} {{Janka}, H.-T. and {Langanke}, K. and {Marek}, A. and {Mart{\'{\i}}nez-Pinedo}, G. and 
%	{M{\"u}ller}, B.}
%\end{thebibliography}

%%%%%%%%%%%%%%%%%%%%%%%%%%%%%%%%%%%%%%%%%%%%%%%%%%

%%%%%%%%%%%%%%%%%%%%%%%%%%%%%%%%%%%%%%%%%%%%%%%%%%

% Don't change these lines
\bsp	% typesetting comment
\label{lastpage}
\end{document}